\documentclass[aps,pre,twocolumn]{revtex4-1}

\usepackage{graphicx}
\usepackage{amsmath}
\usepackage{amssymb}
\usepackage{amsfonts}
\usepackage{color}

\usepackage[utf8]{inputenc}
\usepackage[T1]{fontenc}
\usepackage{lmodern}
\usepackage{floatrow}
\usepackage[normalem]{ulem}
\usepackage{titlesec}

\usepackage{multibib}
\newcites{sel}{Selected works}

\begin{document}

\title{Empirical analysis of vegetation dynamics and the possibility of a catastrophic desertification transition.}

\author{Haim Weissmann$^1$,  Rafi Kent$^2$, Yaron Michael$^2$ and Nadav M. Shnerb$^1$}

\affiliation{$^1$ Department of Physics, Bar-Ilan University, Ramat-Gan
IL52900, Israel. \\ $^2$ Department of Geography and Environment, Bar-Ilan University, Ramat-Gan IL52900, Israel.}


\begin{abstract}
The process of desertification in the semi-arid climatic zone is considered by many as a catastrophic regime shift, since the positive feedback of vegetation density on growth rates yields a system that admits alternative steady states. Some support to this idea comes from the analysis of static patterns, where peaks of the vegetation density histogram were associated with these alternative states. Here we present a large-scale empirical study of vegetation \emph{dynamics}, aimed at identifying and quantifying directly the effects of positive feedback. To do that, we have analyzed vegetation density across  $~2.5 \times 10^6 \  \rm{km}^2$ of the African Sahel region, with spatial resolution of $30 \times 30$ meters, using three consecutive snapshots. The results are mixed. The local vegetation density (measured at a single pixel) moves towards the average of the corresponding rainfall line, indicating a purely negative feedback. On the other hand, the chance of spatial clusters (of many "green" pixels) to expand in the next census is growing with their size, suggesting some positive feedback.  We show that these apparently contradicting results emerge naturally in a model with positive feedback and strong demographic stochasticity, a model that allows for a catastrophic shift only in a certain range of parameters.  Static patterns, like the double peak in the histogram of vegetation density, are shown to vary between censuses, with no apparent correlation with the actual dynamical features.
\end{abstract}

\maketitle

\section{Introduction}

Systems governed by nonlinear dynamics may support alternative steady states \cite{scheffer2001catastrophic}. When such a system is driven by external force it may change its state abruptly (catastrophic shift) at the tipping point, where one of the equilibrium states loses its stability. In ecological systems these shifts are often harmful, causing a loss of bioproductivity and biodiversity, which, in turn, may negatively affect ecosystem functions and  stability. Therefore, the possibility that ecosystems may undergo such an irreversible transition  in response to small and slow environmental variations raises a lot of concern \cite{staal2015synergistic, nes2014tipping,reyer2015forest,baudena2015forests,martinez2016drought,eby2017alternative}.

\begin{figure*}
\includegraphics[width=15cm]{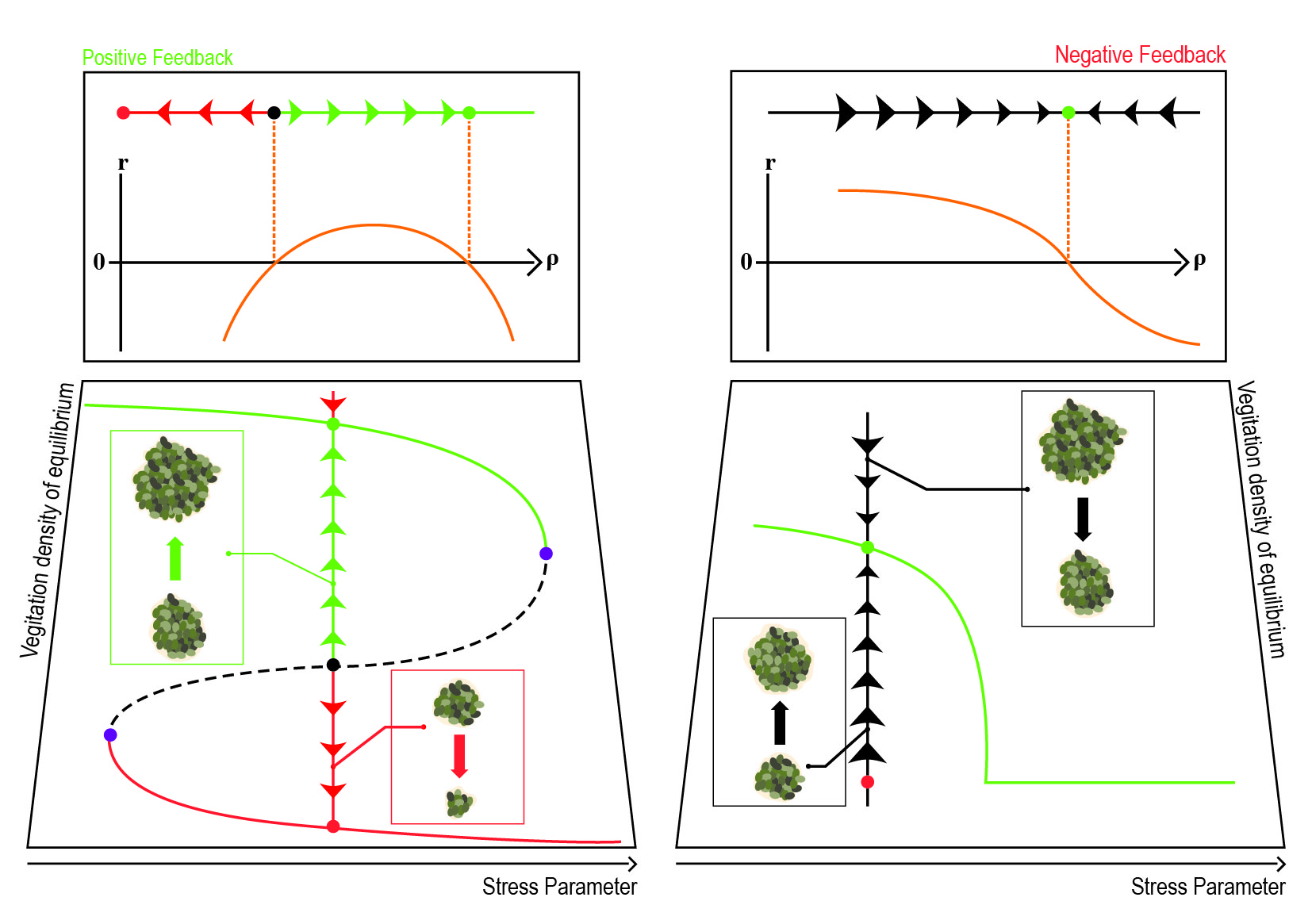}
\caption{\textbf{Systems with positive (left) and  negative (right) feedback}. The right panels depict the continuous transition scenario, where positive feedback is absent or weak.  An increase in the local vegetation density $\rho$ is followed by a decrease in the local growth rate $r$ due to enhanced competition and depletion of resources. Accordingly, a single stable state appears  where $r$ vanishes (upper panel, right). An increase in the stress decreases vegetation density at the equilibrium point (lower panel, right). The chance of a spatial cluster to grow decreases with its size \cite{weissmann2016predicting} and changes sign at the equilibrium point, as illustrated by the arrows.  On spatial domains, the transition (when the density reaches zero) belongs to the directed percolation universality class \cite{kessler2012scaling,bonachela2012patchiness}. The left panels illustrate the case where positive feedback mechanisms allow for local growth only above some critical density. Such a system admits two alternative steady  states (marked green and red),  separated by an unstable fixed point (black). When the stress parameter increases or decreases one of these states may lose its stability at a tipping point (purple) via a saddle-node bifurcation. On spatial domains the transition may be either continuous or discontinuous, depending on the strength of stochasticity \cite{martin2015eluding}. Positive feedback implies that small patches shrink on average, while large patches grow \cite{weissmann2016predicting}.}
 \label{africa}
\end{figure*}

Catastrophic shifts are considered as an important factor in many studies of  transitions between various vegetation regimes, including the destructive process of desertification~\cite{noy1975stability,foley2003regime,janssen2008microscale,sankaran2005determinants}.
Alternative steady states in vegetation systems are the result of a \emph{positive feedback}. In general, a growth of local vegetation density leads to an increase in the competition for limiting resources (like water or sunlight), leading to a decrease of the growth rate. If this effect is dominant, the system does not support alternative states (Figure \ref{africa}, right panels) and the transition is continuous and reversible.  For systems with positive feedback an  increase in the local density leads to an increase in growth rate, hence these systems can support two stable solutions (high density and low density states) for the same set of external parameters (Figure \ref{africa}, left panels). In these systems the  transition may be catastrophic. To explain such a "paradoxical" behavior, an increase of growth rate  despite higher competitive pressure, many  mechanisms (like shading, root augmentation, infiltration rates, fire cycles and so on) have been suggested~\cite{meron2012pattern,gilad2004ecosystem,adams2013mega,tredennick2015effects}.

 However, in spatial systems, positive feedback and alternative steady states of the local dynamics are not sufficient conditions for a catastrophic transition. Even in the presence of these factors, the transition from one stable state to another may be gradual. Two main scenarios of gradual transitions were pointed out in the literature.  First, the effect of stochasticity may lead to a continuous transition, depending on its strength and on system's spatial features \cite{weissmann2014stochastic,martin2015eluding}; second,  local disturbances may generate a moving front between the  two states \cite{bel2012gradual}.

   Recently, an alternative approach to the dynamics of stochastic, spatially extended vegetation systems has been proposed  \cite{weissmann2016predicting}. To this end, one may define a  patch, or a  cluster, as a spatial region where the vegetation density is higher than some threshold, assuming that it corresponds to one of the alternative states (say, the high-density state). In  analogy with the theory of homogenous nucleation in  first order (like ice-water) phase transitions \cite{kelton1991crystal}, the spatial dynamics of vegetation patches reveals the nature of the transition.  This feature manifests itself in the relationship between  the chance of a patch to grow/shrink and its size. If the chance of a patch to expand decreases with its size, the system is controlled by negative spatial feedback, while if this chance increase with size, positive feedback dominates. The regime shift is catastrophic if small patches tend to shrink but large patches tend to grow (see left panels of Figure \ref{africa}). On the other hand, if there is a length scale above which patches tend to shrink, the transition is gradual (see right panels of Figure \ref{africa}).

In  \cite{weissmann2016predicting}, the applicability of this spatial response analysis was demonstrated using simulated data from different generic models. Here we will confront this  approach with ecological reality, using a comparative study of three consecutive states of vegetation in the African Sahel region (south Sahara, see Figure \ref{fig2}A).  Unlike the analyses of static patterns~\cite{hirota2011global,staal2015synergistic,berdugo2017plant}, which assume an underlying dynamical model and try to retrieve its parameters, we (following \cite{dai2013slower}, for example) seek for direct evidence for a dynamics that supports alternative steady states and in particular for its indispensable element, the positive feedback.

\section{Local and spatial response curves: an apparent contradiction}

  We used satellite images of the African Sahel region, obtained at 1999, 2002 and 2015. The EVI index  provides  indications for vegetation density (see Methods for more details) with spatial resolution of $30 \times 30$ meters. Overall our data spans about $2.5 \cdot 10^6 \  km^2$, so the number of elementary pixels is huge and allows for a decent statistical analysis. By comparing the local and spatial  vegetation indices through time, we were able to measure both the local response and the growth/shrink of spatial patches.

\begin{figure*}
\begin{center}
\includegraphics[width=18cm]{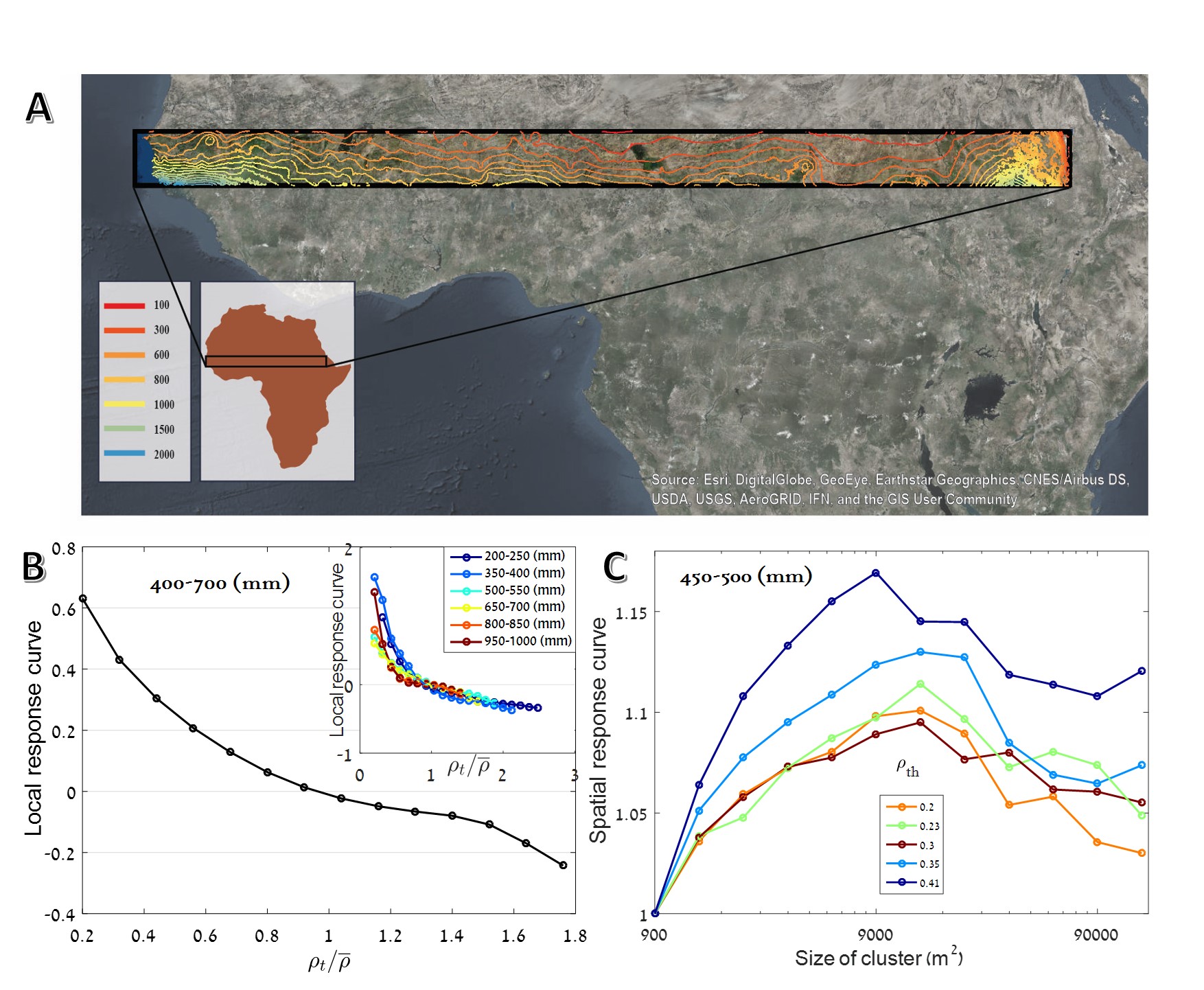}
\vspace{-0.cm}
\end{center}
\caption{\textbf{Temporal analysis of vegetation patterns.} Panel A shows the survey area, the Sahel region in Africa, together with (average) rainfall lines (taken from \cite{hijmans2005very}). In panel B the local response curve, i.e., the  normalized differences between the years 1999 and 2002 ($(\alpha \rho_{2002} - \rho_{1999})/\rho_{1999}$, where $\rho$ is the local vegetation density and the constant $\alpha$ was chosen such that the average growth is zero), is plotted against $\rho_{1999}$ for all pixels between rainfall lines 400-700 $mm/y$ (main plot) and for other precipitation regions (inset), showing what appears to be purely negative feedback (see Supplementary 1.3 for details and errorbars). A spatial cluster is defined as a connected collection of elementary $30 \time 30 m$ squares all, above some threshold vegetation density $\rho_{th}$ (see Methods).  The chance of a cluster of a given spatial area $A$, $P_{\rho_{th}}(A)$, to grow between the years '99 and '02, normalized by the chance of growth of an elementary cluster, $P_{\rho_{th}}(1))$, is plotted in panel C.  Its increase with the size of the cluster, predicted in \cite{weissmann2016predicting}, appear to support the positive feedback hypothesis. Different lines correspond to different threshold density $\rho_{th}$ used to define a "cluster", making it clear that the positive response of small clusters is independent of this definition.}
 \label{fig2}
\end{figure*}

 First we have measured the \emph{local} response of the growth rate to an increase in abundance. To do that, we took for every pixel the local vegetation density, $\rho(t)$, as measured at a census, and compare it with the density at the next census $\rho(t+1)$. The local response is then defined as
\begin{equation}
{\rm Local \ response} \equiv \frac{\alpha \rho(t+1)- \rho(t)}{{\rho(t)}},
\end{equation}
where $\alpha \equiv {\overline \rho(t)}/{\overline \rho(t+1)}$ ($\overline \rho(t)$ is the average density over all pixels at time $t$, or, when we present curves for a certain rainfall line, the average is taken for all pixels in this region)  is a normalization factor. If the response to an increase in vegetation is purely negative (as in the case of logistic or logistic like growth, where due to the increase in vegetation density competition puts more stress on each biomass unit) one expects that a plot of the local response versus $\rho(t)/\overline \rho(t)$ yields a monotonously decreasing curve. The local dynamics supports alternative steady states if, and only if, the local response curve (LRC) first decreases below zero and  then increases above zero as a function of  $\rho(t)/\overline \rho(t)$,  meaning that above some threshold an increase in vegetation density causes the growth rate to take positive values (strong Allee effect).

Empirically, this local response curve was found to be \emph{purely negative}.  For all levels of precipitation and independent of other geographic features, the growth rate of the population decays  with vegetation density, as indicated in Figure \ref{fig2}b (see also Figures S3-S11 for more details and confidence intervals). Sometimes the results resemble a $\theta$-logistic curve with $\theta<1$ (see, however, \cite{clark2010theta}),  but there is almost no indication (except one weak signal in the lowest-left panel of Figure S3) for a local positive feedback. Accordingly, the local response analysis suggests that there is no positive feedback, and hence no alternative steady states, for the semi-arid vegetation system considered here.

Now we would like to implement the technique suggested by \cite{weissmann2016predicting}, i.e., to track the dynamics of spatial clusters of vegetation and to plot their chance to grow against their size; this is what we call the \emph{spatial response curve} (SRC).  A spatial cluster was defined as a collection of adjacent points for which the local density $\rho$ is above some threshold $\rho_{th}$ (different choices of $\rho_{th}$ were examined, see Methods). If this spatial cluster is considered as a "nucleus" of one phase immersed in the background of the other phase, the theory of first order transition suggests that small nuclei shrink in size while large nuclei grow. On the other hand in system that admits only negative feedback the spatial growth of a nucleus decays monotonously with its size.

We have monitored the census to census variation in cluster sizes using the technique implemented in \cite{seri2012neutral,weissmann2016predicting}. Cluster is defined as  a collection of "active" ($\rho > \rho_{th}$) pixels in which every pair is connected by a path of nearest neighbor active pixels.  These clusters were identified in two consecutive censuses, and were associated with each other (i.e., we have decided that a specific cluster at $t+1$ is a modified version of a cluster at $t$) using a motion-detection algorithm. A cluster  of area $A$ ($A$ is the number of pixels in the cluster) at $t$ may shrink, grow, or stay at the same size at $t+1$. We have plotted $P(A)$, the chance an $A$-cluster to grow vs. $A$, in  Figure \ref{fig2}c. To filter out the  overall effect of environmental variations on the density of vegetation (between 1999 and 2002 there is an overall growth of cluster sizes, and between 2002 and 2015 clusters shrink on average), the chance of a cluster of a given size to grow was normalized by the chance of the smallest cluster to grow.

Apparently, Figure \ref{fig2}c and  the supplementary figures S1-S2 indicate that, for any choice of $\rho_{th}$ within reasonable values, there is a clear positive correlation between the patch size and its chance to grow in the next census, at least for small and intermediate size patches.

At first sight, our analysis appears to yield contradictory outcomes. The local (single pixel) response curve shows purely negative feedback (Fig. \ref{fig2}b and figures S3-S11), while the dynamic of vegetation patches with small/intermediate  size (spatial response curve) does indicate positive feedback  (Figs \ref{fig2}c and S1-S2).

\section{Stochastic model as a possible solution}

Interestingly, it turns out that a spatial and stochastic  model, like the one considered recently in \cite{martin2015eluding,weissmann2016predicting}, may (at least in for a certain range of parameters) yield results that will be in agreement with our observations.  Such a model may show purely negative feedback at the local scale, but positive feedback when its cluster dynamics is analyzed.

The model system is the standard Ginzburg-Landau dynamics with diffusion and demographic noise. To be specific, we have simulated the equation,
\begin{equation} \label{eq3}
\frac{\partial \rho(x,y,t)}{\partial t} = D \nabla^2 \rho -a\rho + b\rho^2 -c\rho^3 + {\rm Demographic \  noise},
\end{equation}
where $a$, $b$ and $c$ are positive constants, on a grid of $L \times L$ local sites with periodic boundary conditions. Starting from random initial conditions, the deterministic dynamics (without the noise) is integrated forward in time  (Euler integration with $\Delta t = 0.001$, asynchronous updating). To add demographic noise the biomass of each site, $\rho(x,y,t)$, is replaced by an integer taken from a Poisson distribution with an average $\rho(x,y,t)$ every $\zeta$ elementary timesteps. The smaller $\zeta$ is, the stronger is the demographic noise.

\begin{figure*}
\centerline{\includegraphics[width=15cm]{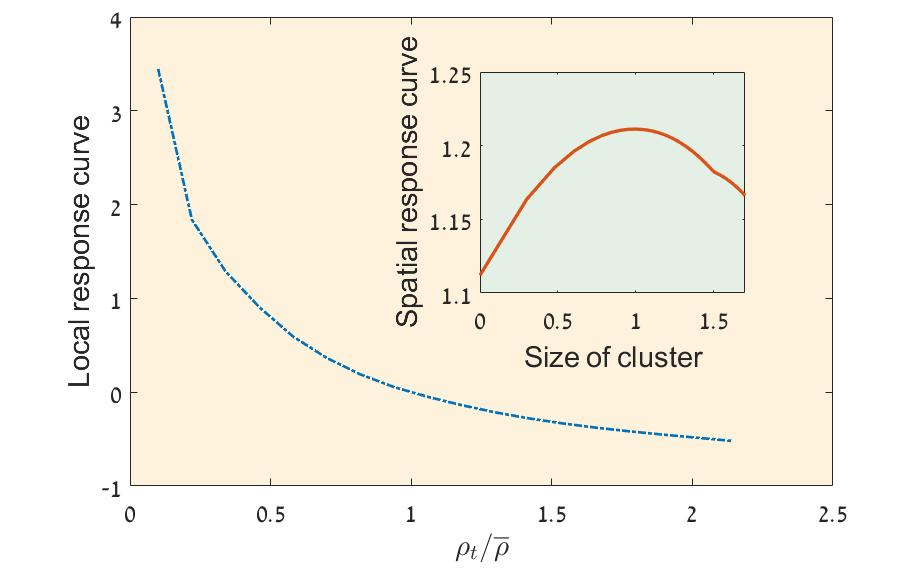}}
\vspace{-0.cm}
\caption{\textbf{The stochastic model.} The local response curve (LRC, main panel) and the spatial response curve (SRC, inset panel) as obtained from simulation of the model analyzed in \cite{martin2015eluding,weissmann2016predicting}. Eq. (\ref{eq3}) was simulated on a $100 \times 100$ lattice with $a=174.5$, $b=40$ $c=1.6$ and $\zeta=2$.  In this parameter regime the deterministic dynamics supports    two alternative steady states (one of them absorbing) and the demographic stochasticity is relatively strong.   The LRC is negative, while the SRC is positive, in agreement with the empirical result presented above.}
 \label{fig4}
\end{figure*}

As demonstrated in Fig.  \ref{fig4}, this simple model, when simulated with relatively strong demographic noise, yields, in fact, purely negative local response and positive spatial response curve at the same time. Although we cannot explain the mechanism behind this behavior, at least it provides a possible framework for further analysis.

 As explained by \cite{martin2015eluding}, the model (\ref{eq3}) allows for two types of transitions: when the demographic noise is strong the transition is continuous and reversible, with no tipping points and catastrophic events, while for weak noise the transition is (as in the purely deterministic case) irreversible and catastrophic.  We do not know, yet, in what parameter region of this model one finds  the behavior demonstrated in Fig. \ref{fig4}, i.e., negative local response and positive spatial response, but it seems  more likely  that this behavior is a characteristic of the strong noise, continuous transition regime.  Clearly, more work is needed in order  to support the interpretation of our result using this specific model and  to clarify the connection between the local/spatial response and the different phase transitions in this model.

\section{The irrelevance of vegetation cover distributions}

  During the last decade many researchers have used features of a single snapshot of  spatial patterns, like the tree-cover histogram or the cluster statistics, as indicators for the state of the system. They assumed an underlying simple two-state model and retrieved its parameters from the static patterns, a procedure that allows them to assess the resilience of each of the states and to predict its response to environmental variations~\cite{hirota2011global,staal2015synergistic,berdugo2017plant}. In particular, a double peak in the vegetation density histogram is considered as an indication for two alternative steady state, and its shape reflects the weights of the attractors.  These snapshot analysis approaches were criticized by some authors \cite{ratajczak2012comment,ratajczak2014abrupt,staver2011global,hanan2014analysis}, but our data allow, for the first time,  to examine the relevance of the static features to the actual time evolution of the system.

\begin{figure*}
\begin{center}
\includegraphics[width=18cm]{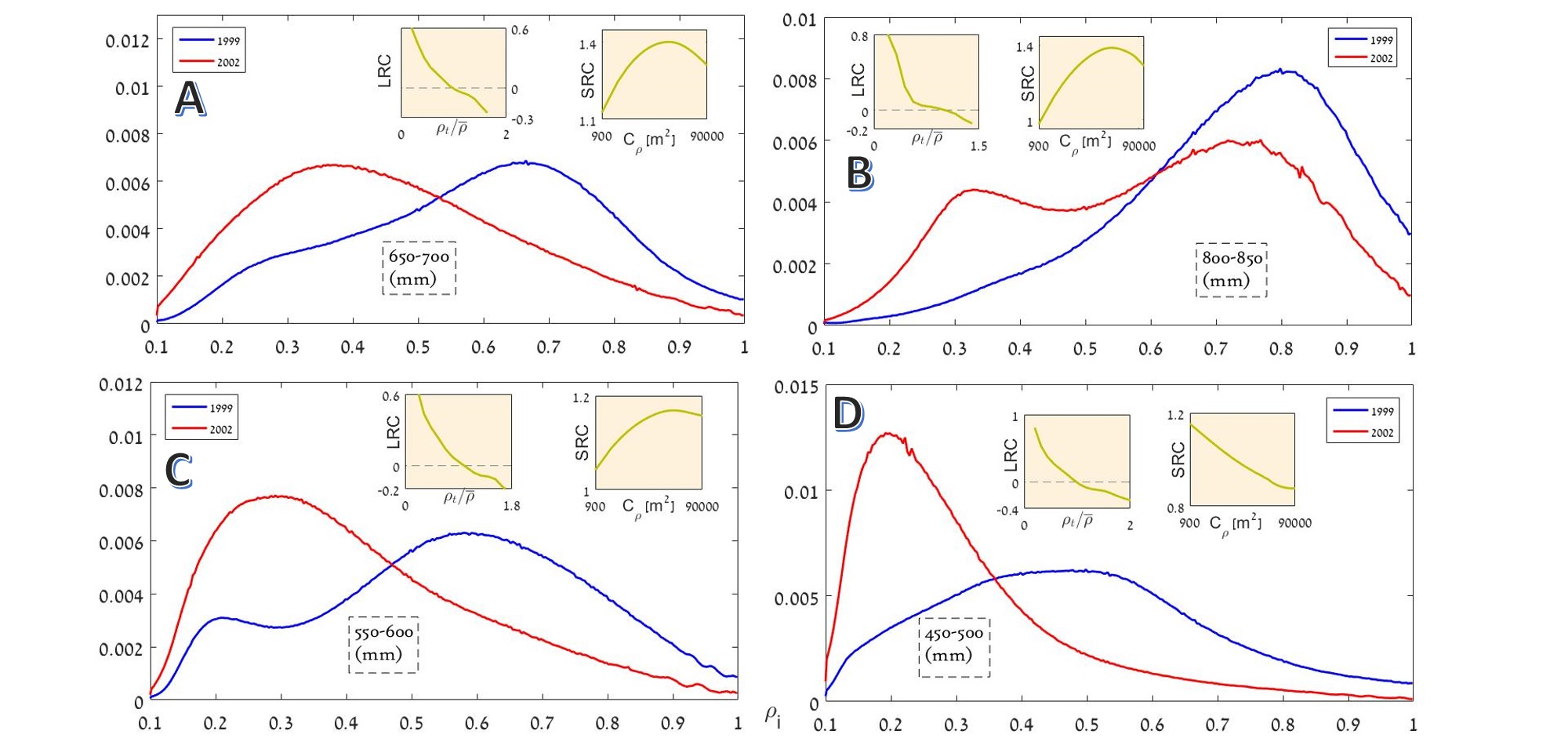}
\vspace{-0.cm}
\end{center}
\caption{ The relative frequency distribution of vegetation cover is plotted here for four levels of mean annual precipitation using the data of 1999 (blue) and 2002 (red). The average vegetation cover decreased during this period, so the histograms of 2002 are shifted systematically to the left. The histograms for 650-700 mm/y (panel A) admit a clear single  peak for both years, in the region 800-850 mm/y (panel B) one observes a crossover from a unimodal to bimodal distribution and for 550-600 mm/y (panel C) the histogram has a double peak in 1999 and a single peak in 2002, meaning that the modality of the histogram is \emph{not} a robust feature of the system.  The local response curve (LRC) in all cases has a negative slope (left inset). The spatial response curve (SRC) shows signs of positive feedback in most of the cases, but there are some rare exceptions, one is demonstrated in panel D. See supplementary figures S2-S12 for more details.}
 \label{fig3}
\end{figure*}

 We obtained disappointing results.  The main static features considered so far,  the histogram of vegetation densities, appears to be irrelevant  to the dynamics of the system. As demonstrated in Figure \ref{fig3} (see also figure S12), histograms of vegetation densities along equi-precipitation lines may admit a single peak in one census and a double peak three years later. In any case there is no apparent correlation between the structure of the histogram and the level of positive/negative feedback observed in cluster analysis.

\section{Discussion}

 Obviously, the outcome of spatio-temporal analyses may depend on the spatial and temporal scales considered. The $30 \times 30 m$ resolution imposed by the remote sensing limitations and the $3-15$ years  between snapshots put some constraints on our ability to examine the vegetation system dynamics. Still, we believe that our temporal resolution is appropriate: the correlations between vegetation density on a single pixel level is about $50\%$, meaning that we observed substantial variations yet the system did not forget its initial state (see table S1 for more details). We carried out the same analysis on coarser scales, from $60 \times 60m$ to $480 \times 480 m$, and found again negative local feedback and a collection of single and double peak vegetation density histograms without any apparent correlation with dynamical response curves.

  We conclude that, unlike static patterns, the response curves that characterize the dynamics appear to be robust and to indicate purely negative local response  and positive spatial response for small-to-intermediate patches.   We believe that our results  demonstrate the importance of noise and spatial structure in this system, and the necessity to interpret the results using models that admit these features. When noise and spatial structure are taken into account  even a system with positive feedback may support a continuous transition instead of a catastrophic shift, and one cannot ignore the possibility that this is indeed  the case in the Sahel region.

\section{Methods}
\subsection{Vegetation index} Through this paper we have used remote-sensing data to estimate the vegetation density in the Sahel region.  The most commonly used vegetation index is  NDVI (normalized difference vegetation index). However, this index  is sensitive to background soil reflectance as well as atmospheric disturbances. To correct for these effects, here we used another index,  EVI (enhanced vegetation index), that incorporates  corrections to both soil reflectance and atmospheric disturbances. While it is mostly used in high production areas (e.g., tropical forests) we find it appropriate to discriminate between light colored soil and vegetation a very common scenario in our study area \cite{wallace2008estimation,huete2002overview}.

To produce EVI layers, we  downloaded, using the Google Earth Engine platform \cite{googleearthengine}, Landsat 7 images from the studied area (latitude: $11.7^o - 15.1^o$ and longitude: $17.8^o - 40^o$). We  used records between $15/09/1999-15/10/1999$ and the same dates in 2002 and 2015. To exclude images with high cloud cover, as well as to overcome a known $22\%$ data loss in 2015 \cite{landset}, we applied the greenest pixel composite filter \cite{shresthamulti} in which the pixel that has the highest green reflectivity value was selected from the images available within the date range.

\subsection{Defining $\rho_{th}$} The theory of homogenous nucleation predicts that a "grain" of the preferred phase (say, a grain of ice inside water at a temperature below $0^o \  C$) will grow if its spatial size is large and will shrink if it is too small. In this paper we have monitored this behavior, plotting the chance of a local patch to grow as a function of its size. However, unlike the water-ice scenario, we do not have a well defined criteria that identify the different "phases". Observing a collection of ten pixels with a certain EVI index, say, one may still ask if they belong to the high vegetation or to the low vegetation phase.

In principle, the answer to this question is to plot a histogram of vegetation cover and to identify the two peaks with the different phases. However, as seen in the above (figure 3)
and in figure S12, in many cases this histogram does not show a bimodal behavior.

To overcome this difficulty, we  used the following procedure: when the vegetation density histogram has a double peak, $\rho_{th}$ was taken to be the one that corresponds to the maximum deep between these two peaks (the point that corresponds to the unstable fixed point separating the two metastable phases, as one expects in the classical theory of phase transitions). When we found only a single peak, $\rho_{th}$ was taken at the peak itself (as one expects if a strong noise blurs the details of the transition). To be on the safe side we tried all kind of different $\rho_{th}$ values. Figure 2c  shows that the general trends are independent of that choice.

\subsection{Clusters tracking} To track the evolution of clusters we  implemented a simple motion tracking algorithm (see, e.g., \cite{seri2012neutral,falkowski2006mining,hartmann2014clustering}). Each cluster at one snapshot is compared with the previous one to identify growth or decay, the details of this analysis are given in \cite{weissmann2016predicting}.

\section{\bf Acknowledgments} We acknowledge the support of the Israel
Science Foundation, grant no. $1427/15$.

\bibliography{Paper_Data_africa_references}
\bibliographystyle{pnas2011}


\onecolumngrid
\pagebreak

\setcounter{equation}{0}
\setcounter{section}{0}
\setcounter{figure}{0}
\setcounter{table}{0}
\setcounter{page}{1}
\makeatletter
\renewcommand{\theequation}{S\arabic{equation}}
\renewcommand{\thefigure}{S\arabic{figure}}

\begin{center}
\Large{\bf{Supplementary information to: Empirical analysis of vegetation dynamics and the possibility of a catastrophic desertification transition}}
\end{center}

\section*{1. More results}
\subsection*{1.1 Correlation coefficient and number of pixels} Table S1 shows the number of pixels for each set of rainfall lines and the correlation coefficient between the years 1999-2002 and 2002-2015. In general the correlation coefficients are between 0.3 and 0.7, meaning that there is a substantial overlap between the local measures of vegetation cover, still there are also substantial modifications so the tracking of clusters and local cover is meaningful.

\begin{table}[H]
    \includegraphics[scale = 0.6]{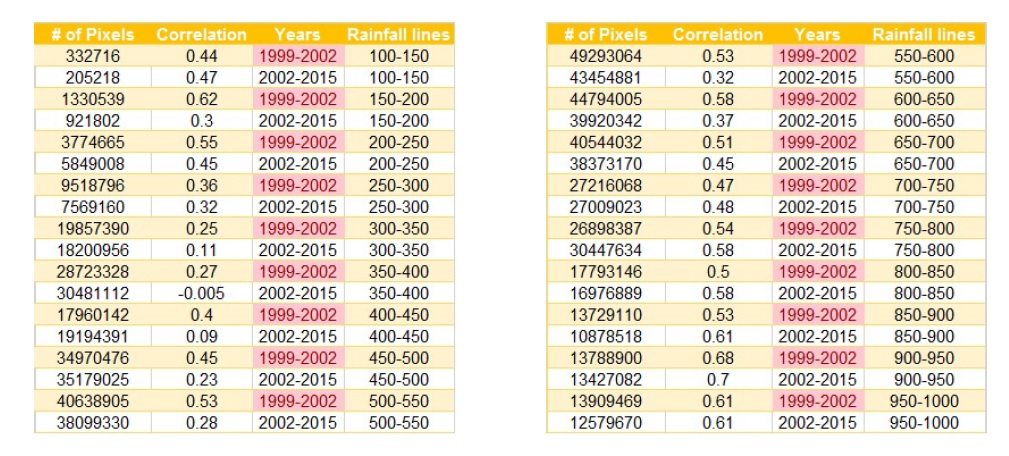}
    \caption{Correlation coefficient and number of pixels for rainfall lines 100-1000 [$mm/year$].}
\end{table}


\subsection*{1.2 Spatial response curve}
 The figures below, S1 and S2, provide supplemental information to the data presented in Fig. 2C of the main paper, that refers only to the region between 450 and 500 mm/y. Here the results are  presented for all rainfall lines. For each rainfall regime, the blue open circles are the data, i.e., the average growth/shrink in size of a cluster of $n$ pixels ($\log_{10}(n)$ is the x-axis).  To show the general trend we have smoothed the data using the Matlab smooth algorithm (black full line, span parameter: 0.9, method: loess) \citesel{matlab}. In most cases one sees a clear positive feedback for small clusters but for large clusters the line curves down. A fit of the data to a parabola (full red line) using the function "glm" in R environment \citesel{R_environment} is very similar to the smooth curve (the Student t-test with the data gives (unless otherwise noted) p-values below $5 \%$)  in most cases and indicates better the decay in the growth rate for large clusters. Green lines appear when the p-value for a parabola was too big and we have used linear regression instead.

\newpage
\pagebreak

\begin{figure}[H]
	 \makebox[\textwidth][c]{\includegraphics[scale=0.35]{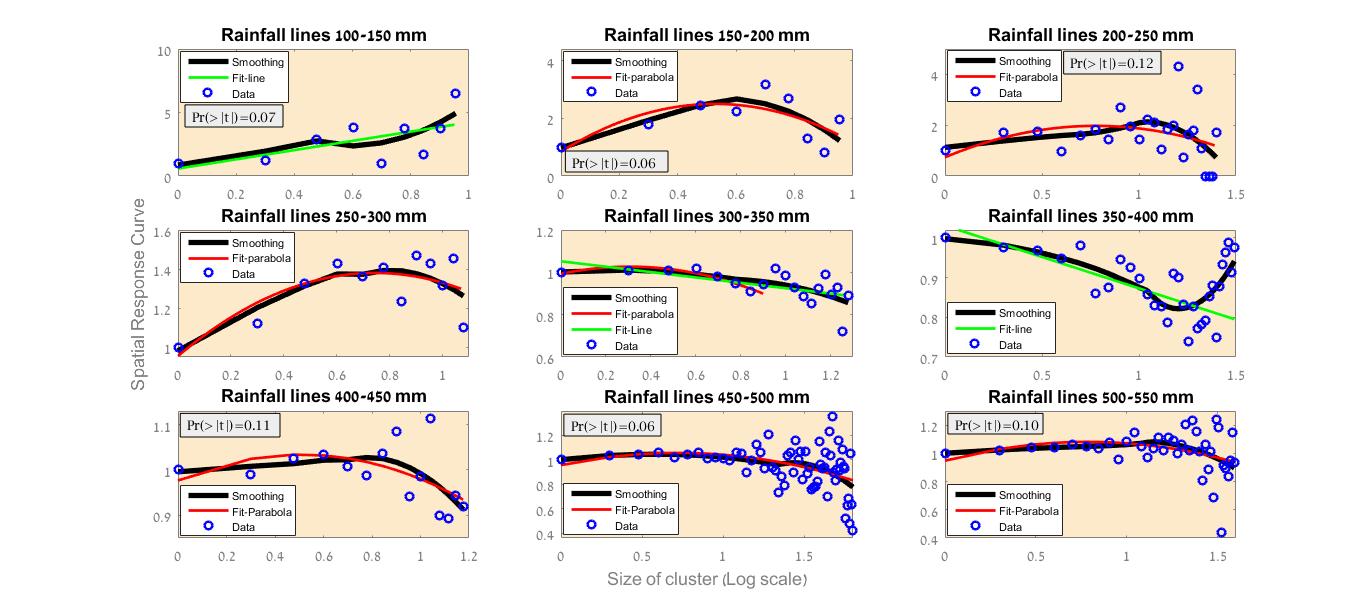}}

	\vspace{1.0pt}
	
	 \makebox[\textwidth][c]{\includegraphics[scale=0.35]{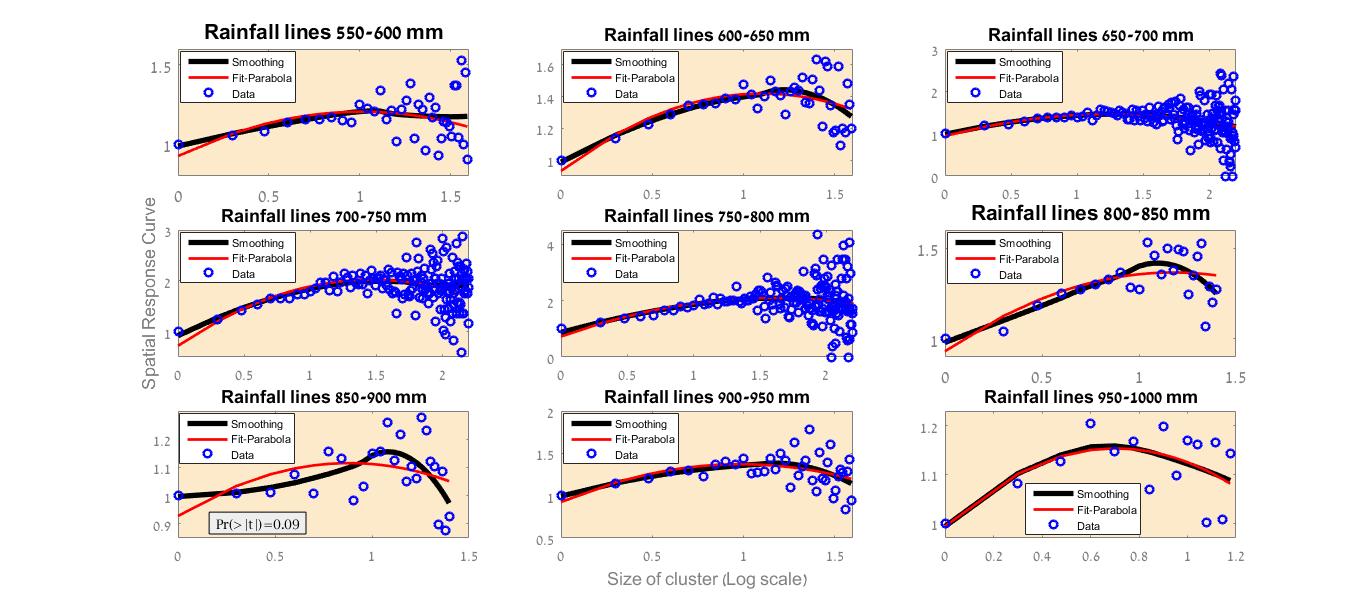}}
	\caption{\textbf{Spatial response curve between years 1999-2002 (rainfall lines 100 - 1000 [$mm/year$])}.}
\end{figure}

\begin{figure}[H]
	 \makebox[\textwidth][c]{\includegraphics[scale=0.35]{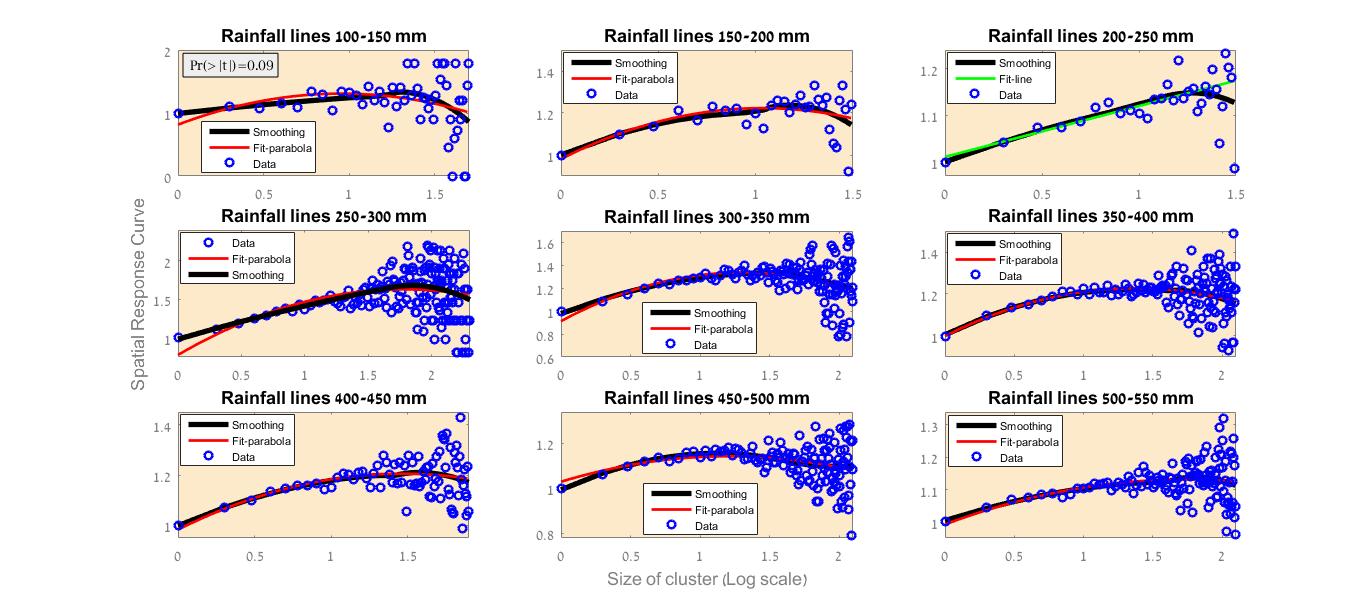}}

	\vspace{1.0pt}
	
	 \makebox[\textwidth][c]{\includegraphics[scale=0.35]{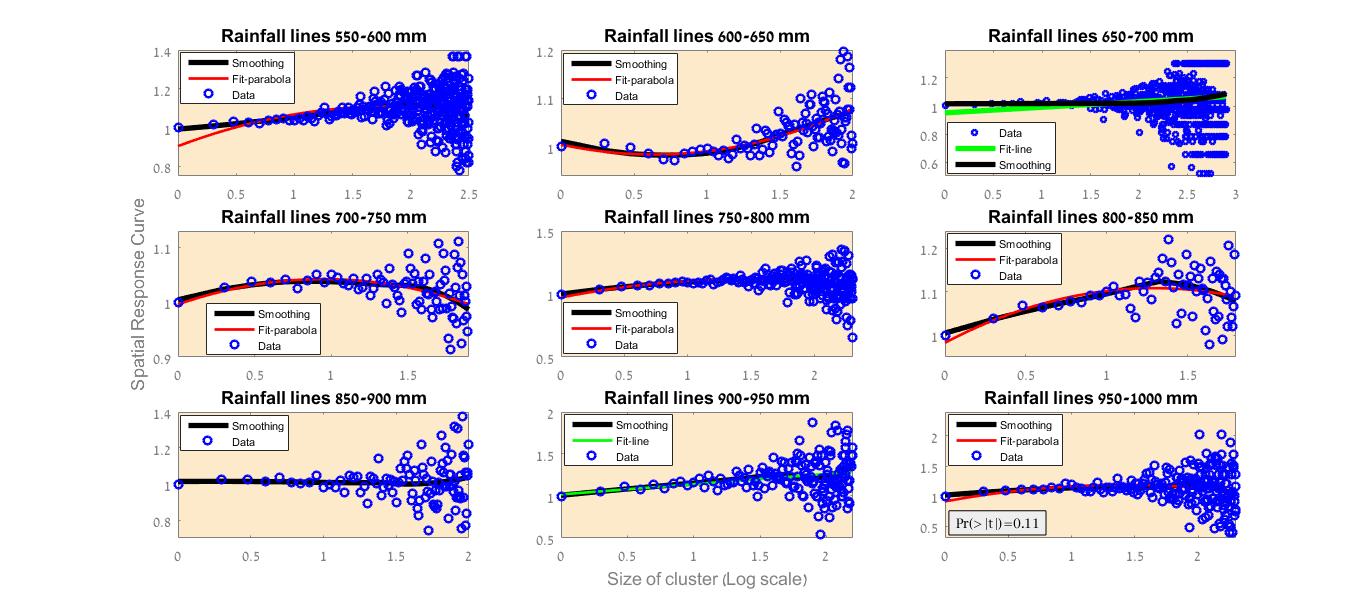}}
	\caption{\textbf{Spatial response curve between years 2002-2015 (rainfall lines 100 - 1000 [$mm/year$])}.}
\end{figure}

\pagebreak
\subsection*{1.3 Local response curve} The figures below,  S3-S11, is a supplement for Fig. 2B of the main text. The local response is presented for different rainfall lines.

Given the vegetation density at a pixel, $\rho_{ij}^t$ (in 1999, say) and the density at the same pixel in the next census (say, 2002), $\rho_{ij}^{t+1}$,  the local response is defined as:
\begin{eqnarray} \label{LRC}
LRC_{ij}=\frac{\alpha \rho_{ij}^{t+1}-\rho_{ij}^t}{\rho_{ij}^t}
\end{eqnarray}
where  $$\alpha \equiv \frac{\langle \rho^t \rangle}{\langle \rho^{t+1} \rangle},$$
 and $\langle \rho^t \rangle$ is the average (over all the rainfall line area) vegetation density in the first census.

In the panels below we present $LRC$ vs.  $\rho_{ij}^t/\langle \rho^t \rangle$ (errorbars in S4-S5 with $STD=1$). The response is clearly negative, except for the years 99-02 for rainlines 850-900, where one may notice weak positive feedback.

\newpage
\pagebreak

\begin{figure}[H]
	 \makebox[\textwidth][c]{\includegraphics[scale=0.5]{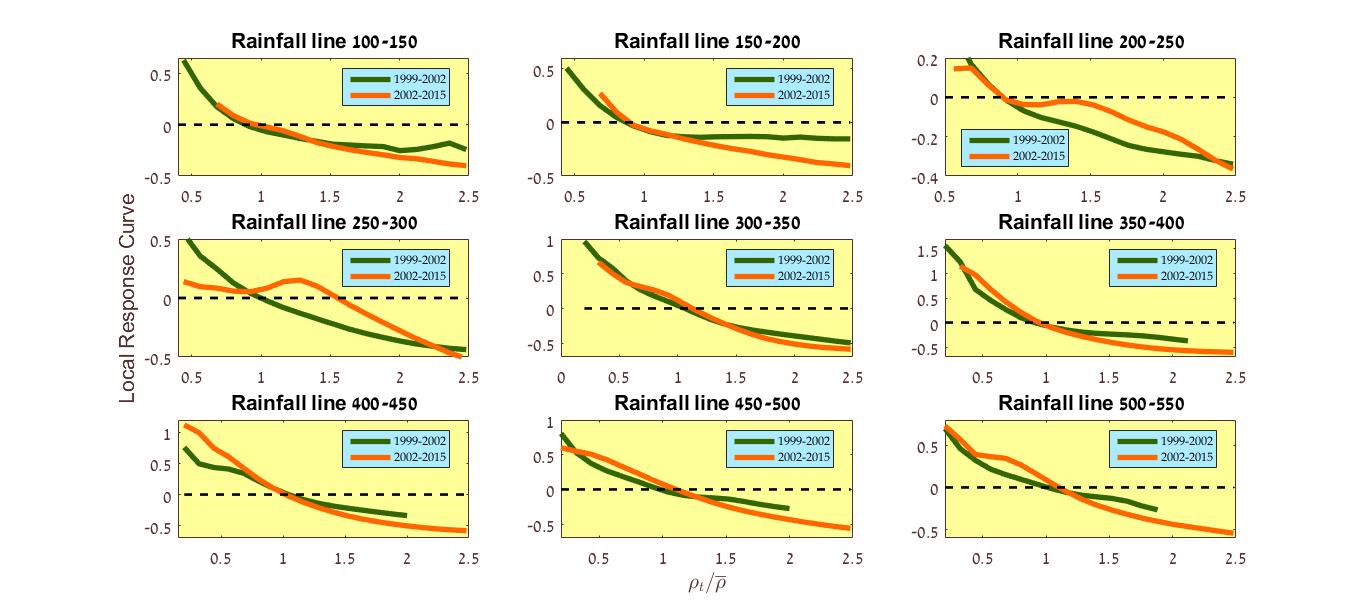}}

	\vspace{1.0pt}
	
	 \makebox[\textwidth][c]{\includegraphics[scale=0.5]{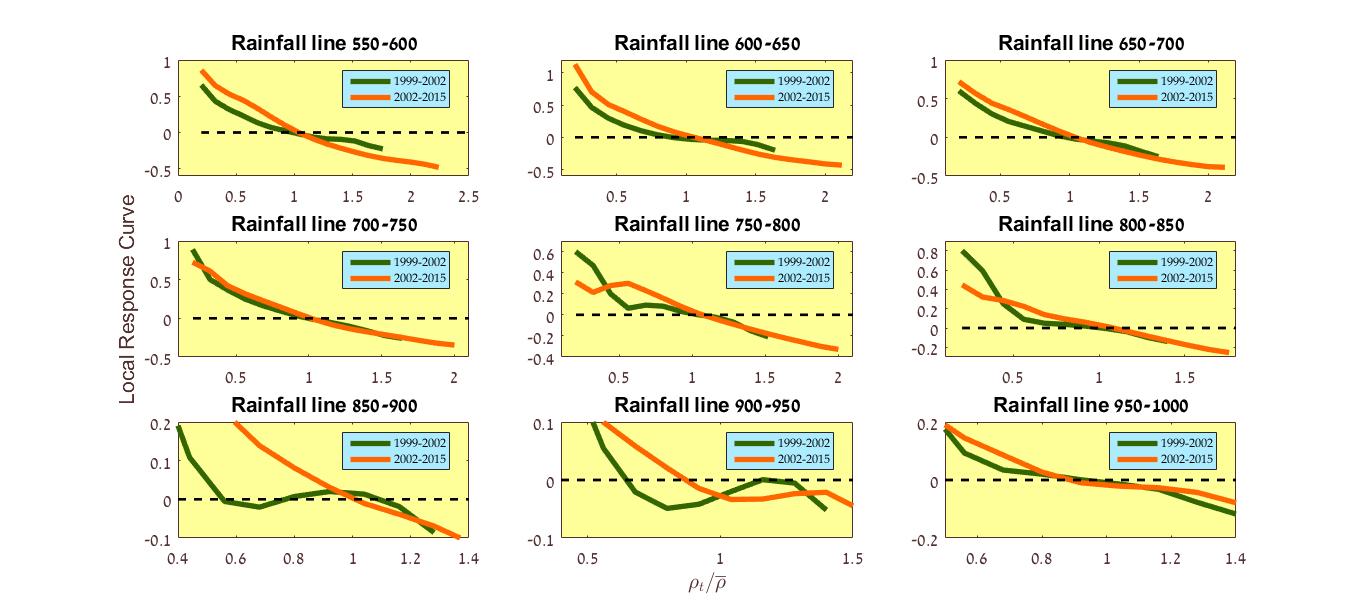}}
	\caption{\textbf{Local response curve (rainfall lines 100 - 1000 [$mm/year$])}.}
\end{figure}

\begin{figure}[H]
	 \makebox[\textwidth][c]{\includegraphics[scale=0.5]{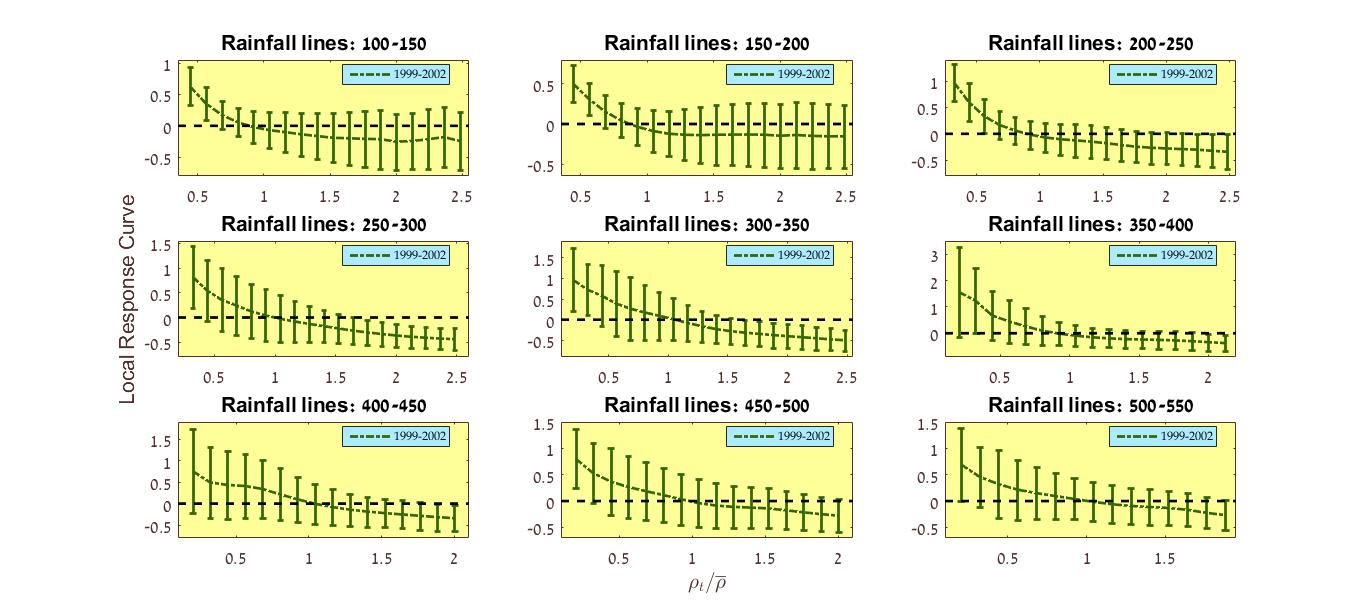}}

	\vspace{1.0pt}
	
	 \makebox[\textwidth][c]{\includegraphics[scale=0.5]{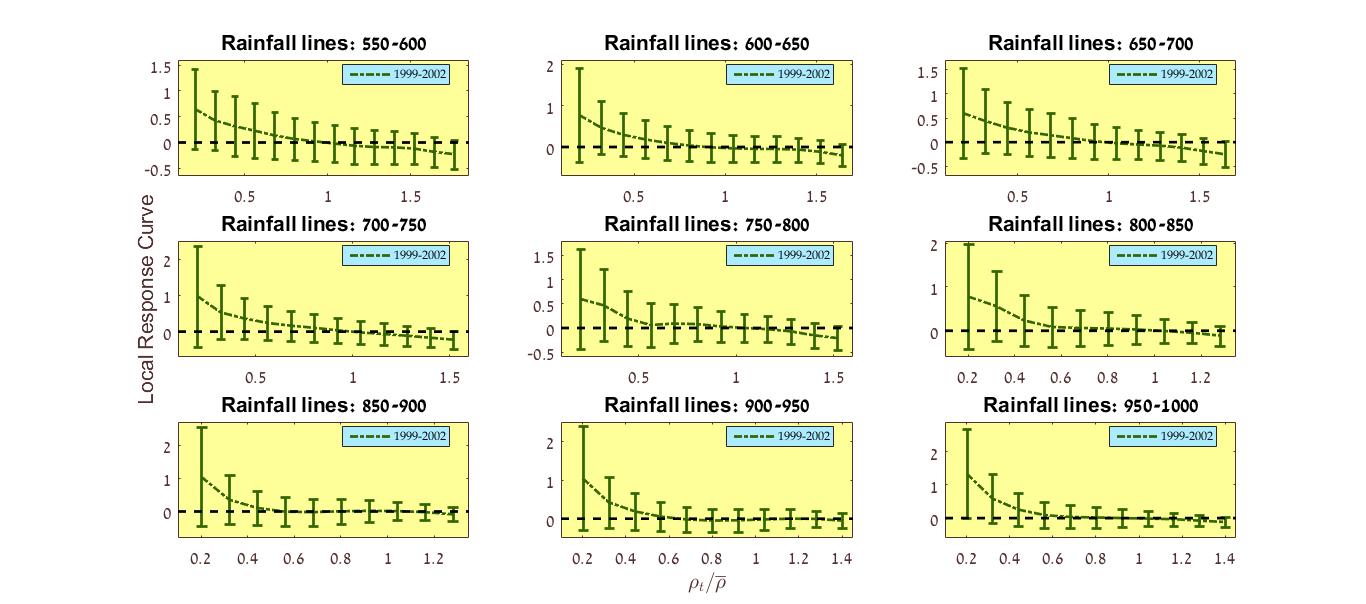}}
	\caption{\textbf{Local response curve - with errorbars (rainfall lines 100 - 1000 [$mm/year$], years 1999-2002)}.}
\end{figure}

\begin{figure}[H]
	 \makebox[\textwidth][c]{\includegraphics[scale=0.5]{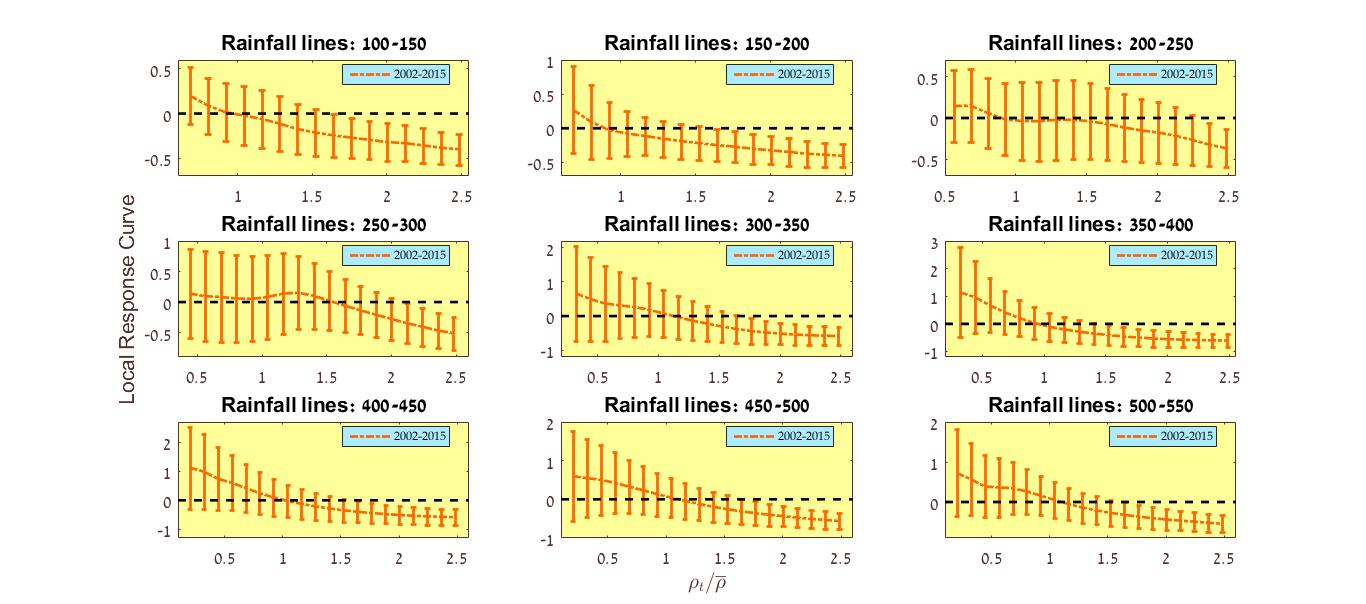}}

	\vspace{1.0pt}
	
	 \makebox[\textwidth][c]{\includegraphics[scale=0.5]{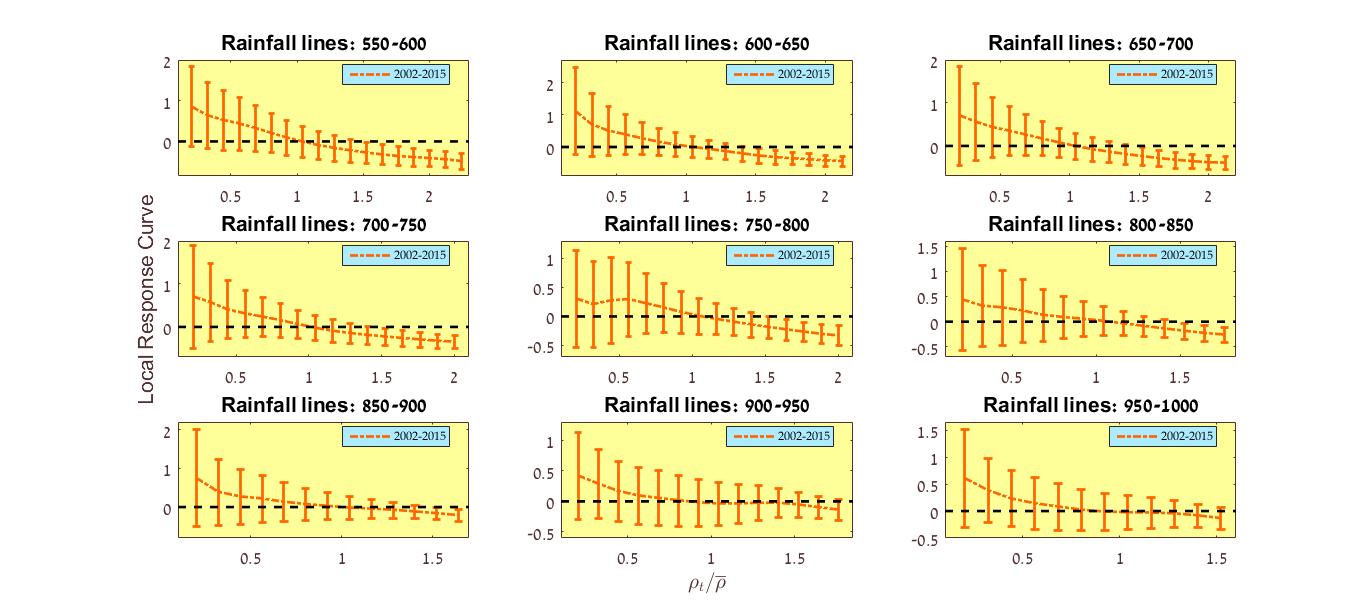}}
	\caption{\textbf{Local response curve - with errorbars (rainfall lines 100 - 1000 [$mm/year$], years 2002-2015)}.}
\end{figure}

\begin{figure}[H]
	 \makebox[\textwidth][c]{\includegraphics[scale=0.5]{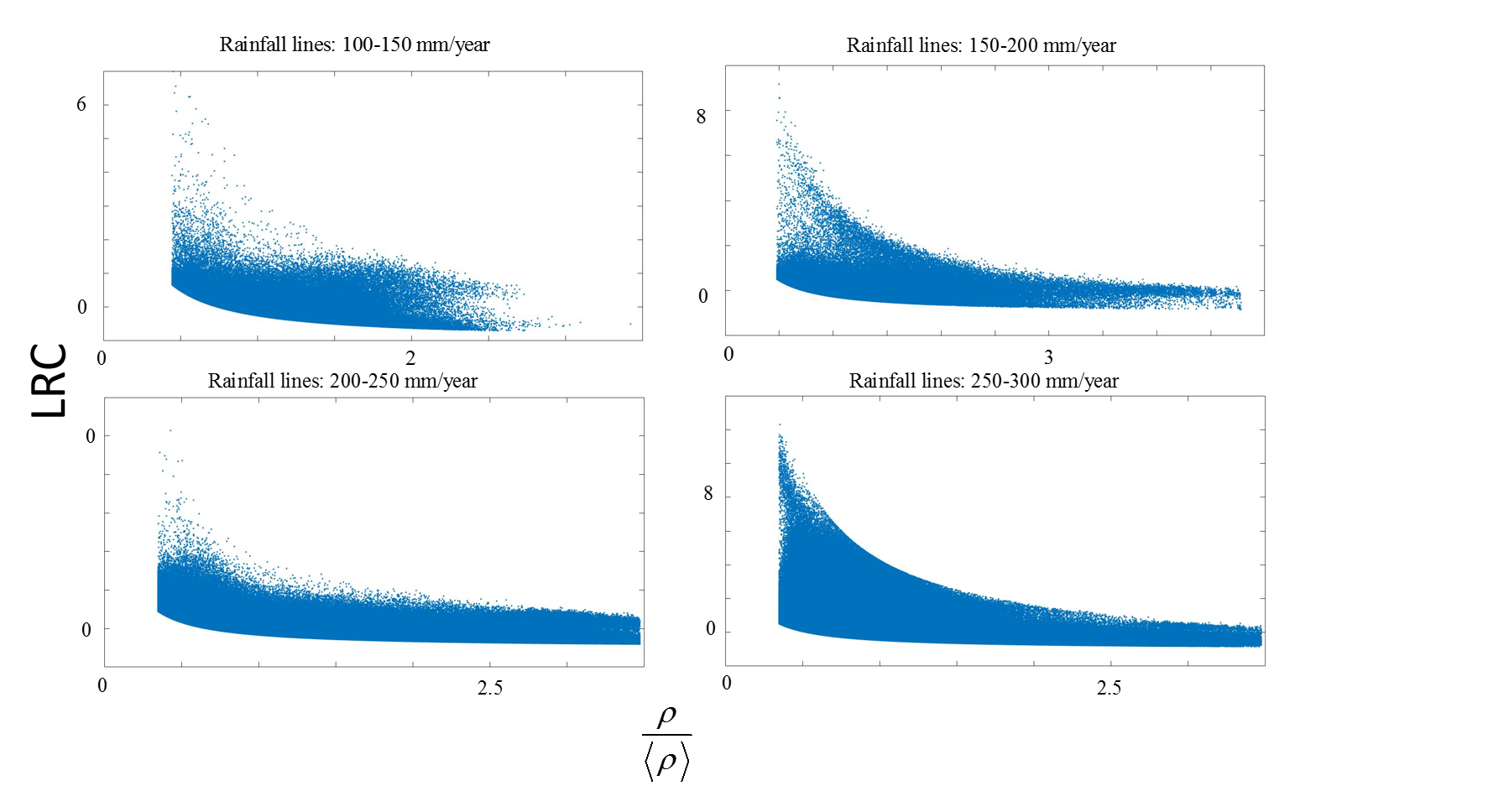}}

	\vspace{1.0pt}
	
	 \makebox[\textwidth][c]{\includegraphics[scale=0.5]{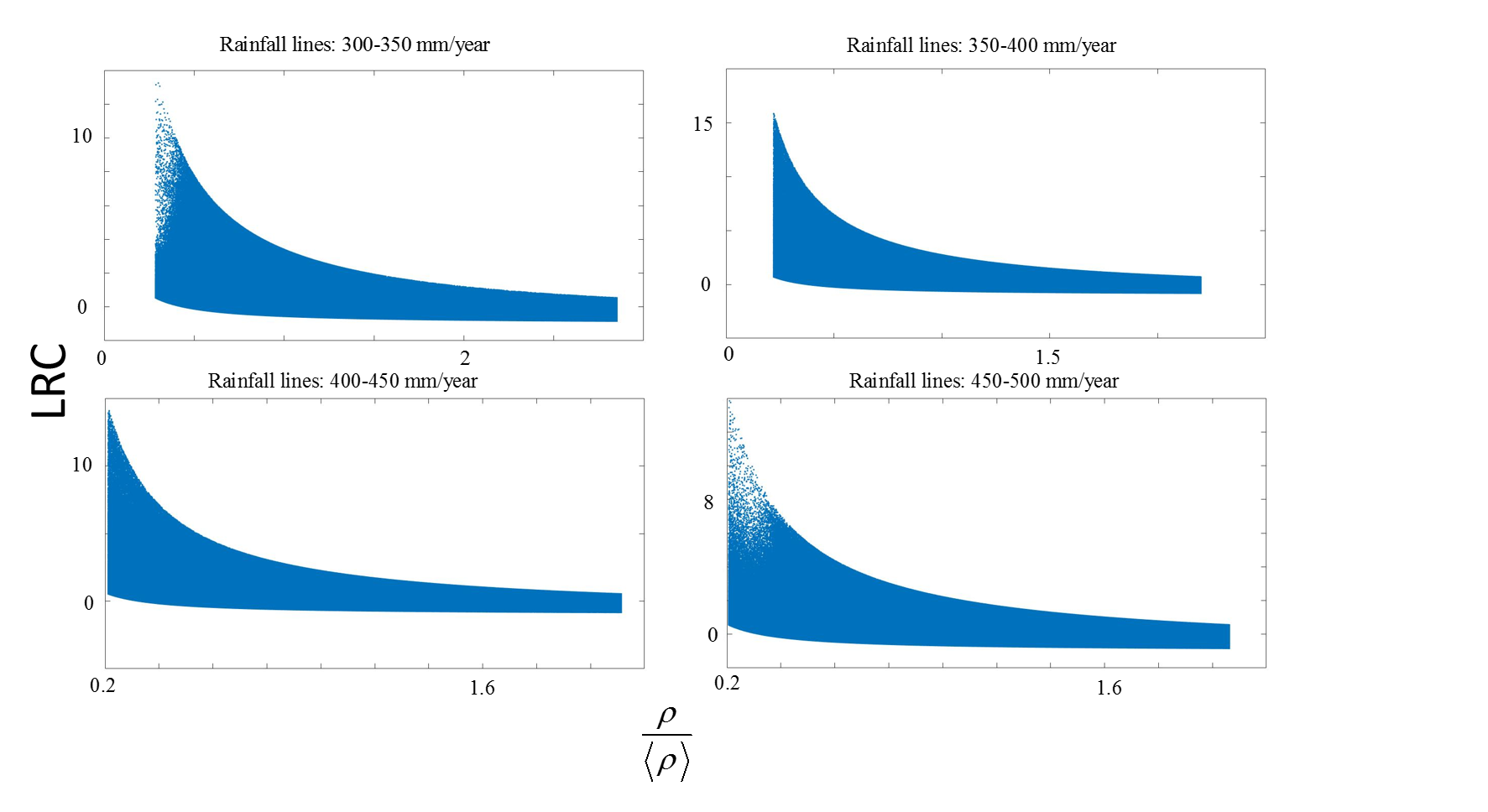}}
	\caption{\textbf{Local response curve - Data points (rainfall lines 100 - 500 [$mm/year$], years 1999-2002)}.}
\end{figure}

\begin{figure}[H]
	 \makebox[\textwidth][c]{\includegraphics[scale=0.5]{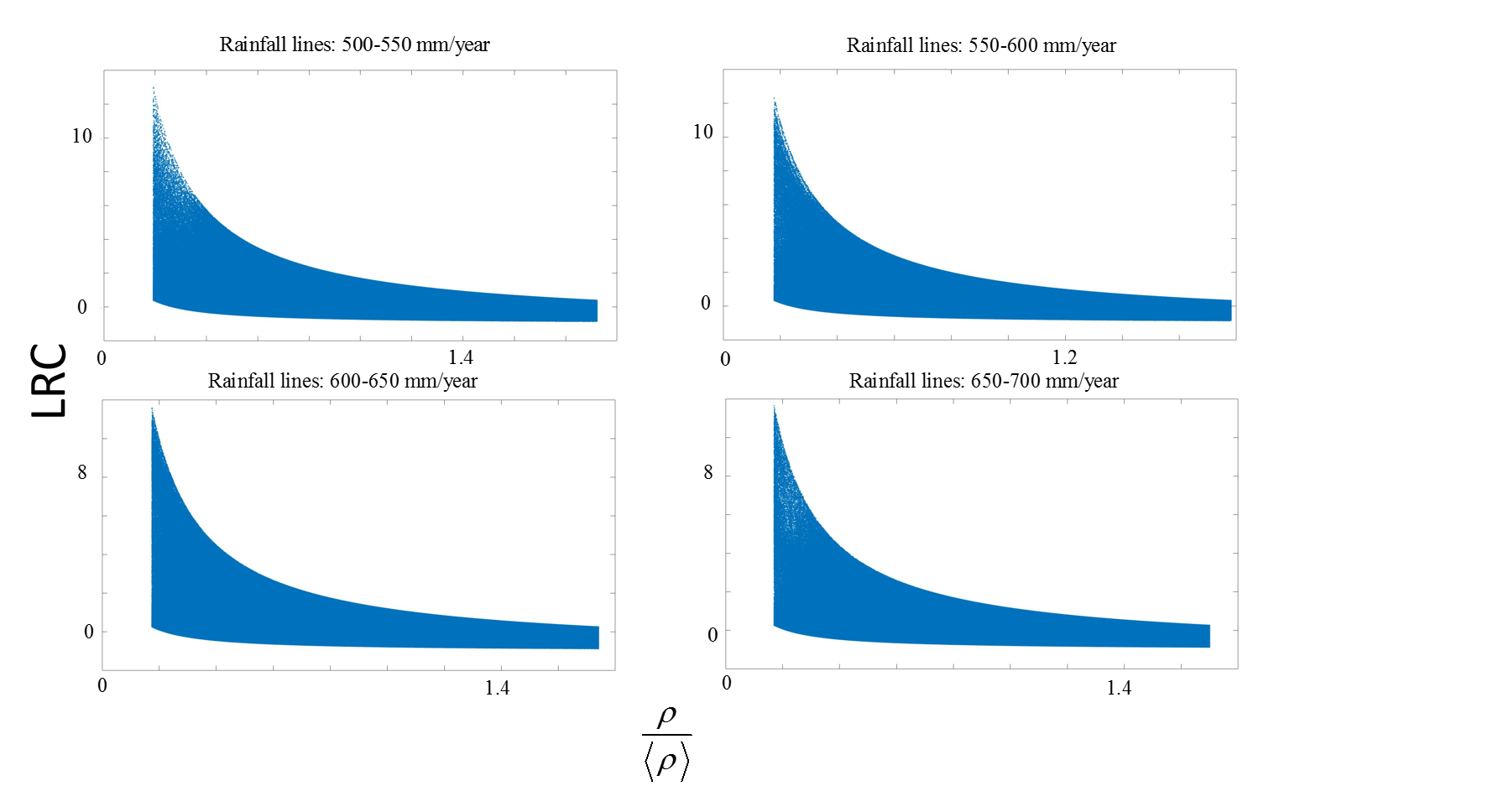}}

	\vspace{1.0pt}
	
	 \makebox[\textwidth][c]{\includegraphics[scale=0.5]{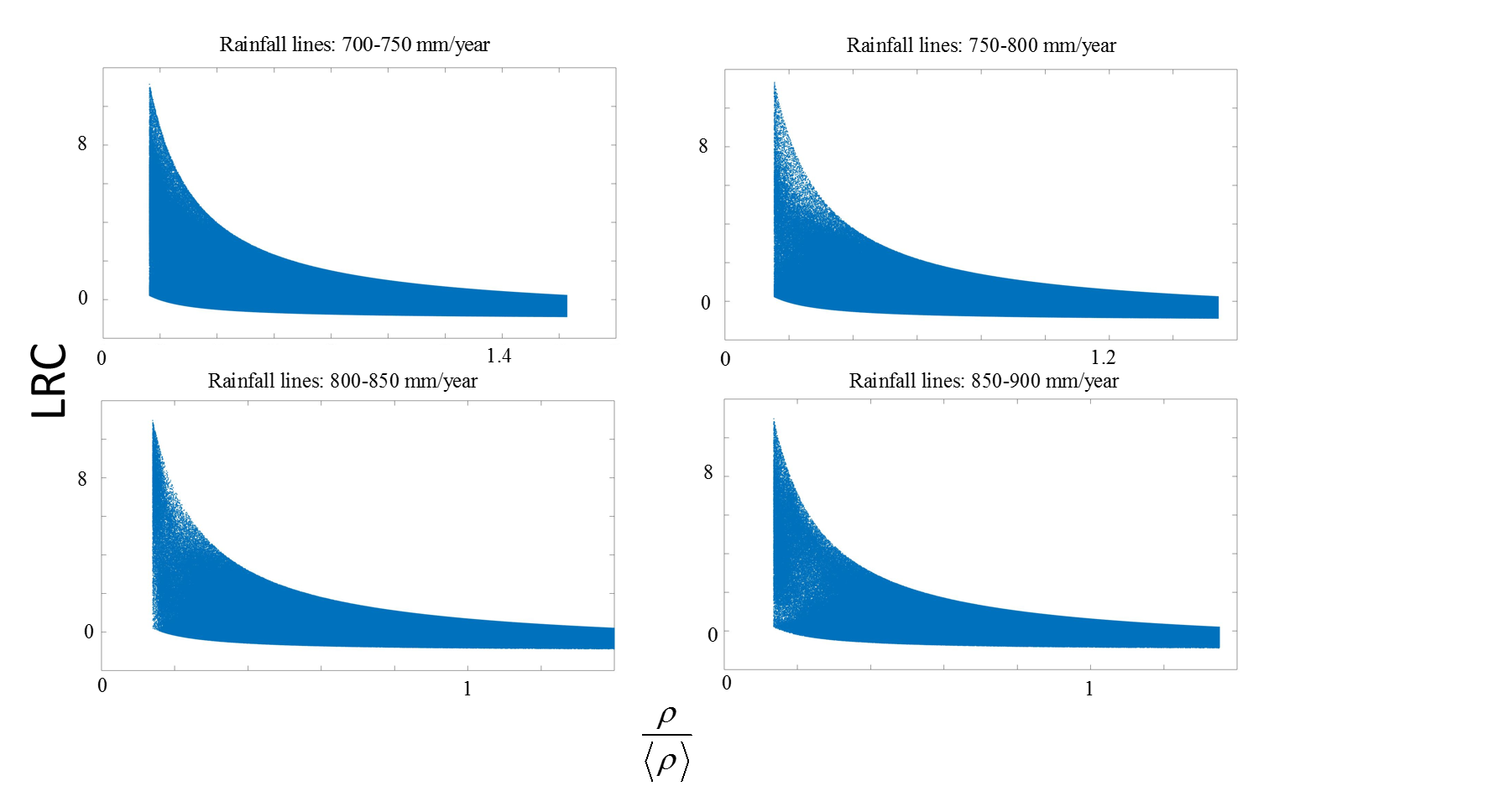}}
	\caption{\textbf{Local response curve - Data points (rainfall lines 500 - 900 [$mm/year$], years 1999-2002)}.}
\end{figure}

\newpage
\pagebreak

\begin{figure}[H]
	 \makebox[\textwidth][c]{\includegraphics[scale=0.5]{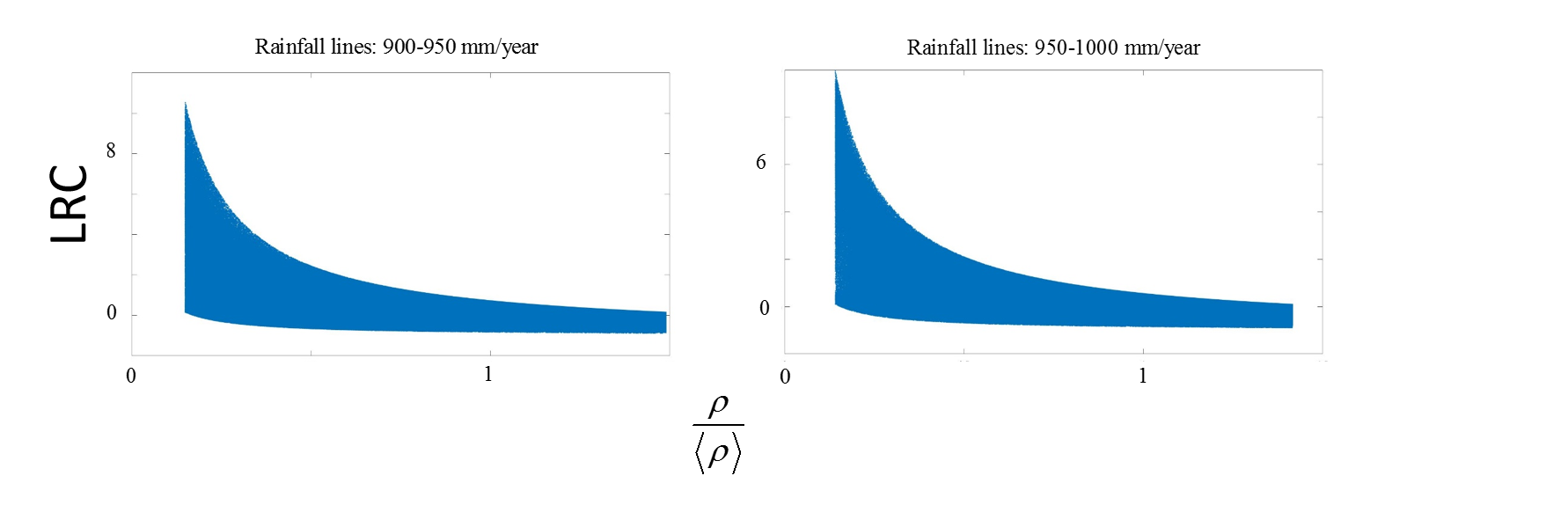}}

	\caption{\textbf{Local response curve - Data points (rainfall lines 900 - 1000 [$mm/year$], years 1999-2002)}.}
\end{figure}

\newpage
\pagebreak

\begin{figure}[H]
	 \makebox[\textwidth][c]{\includegraphics[scale=0.5]{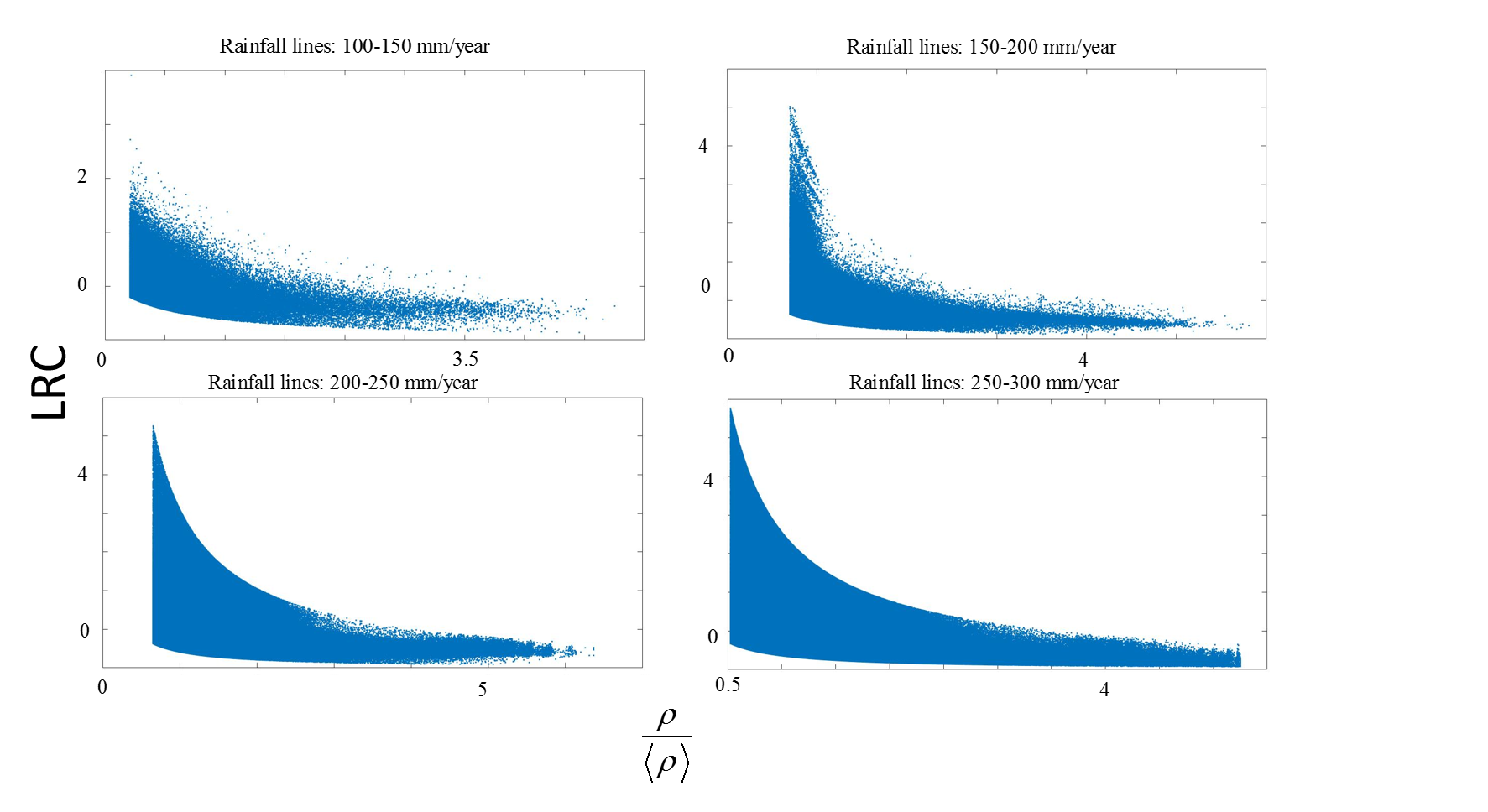}}

	\vspace{1.0pt}
	
	 \makebox[\textwidth][c]{\includegraphics[scale=0.5]{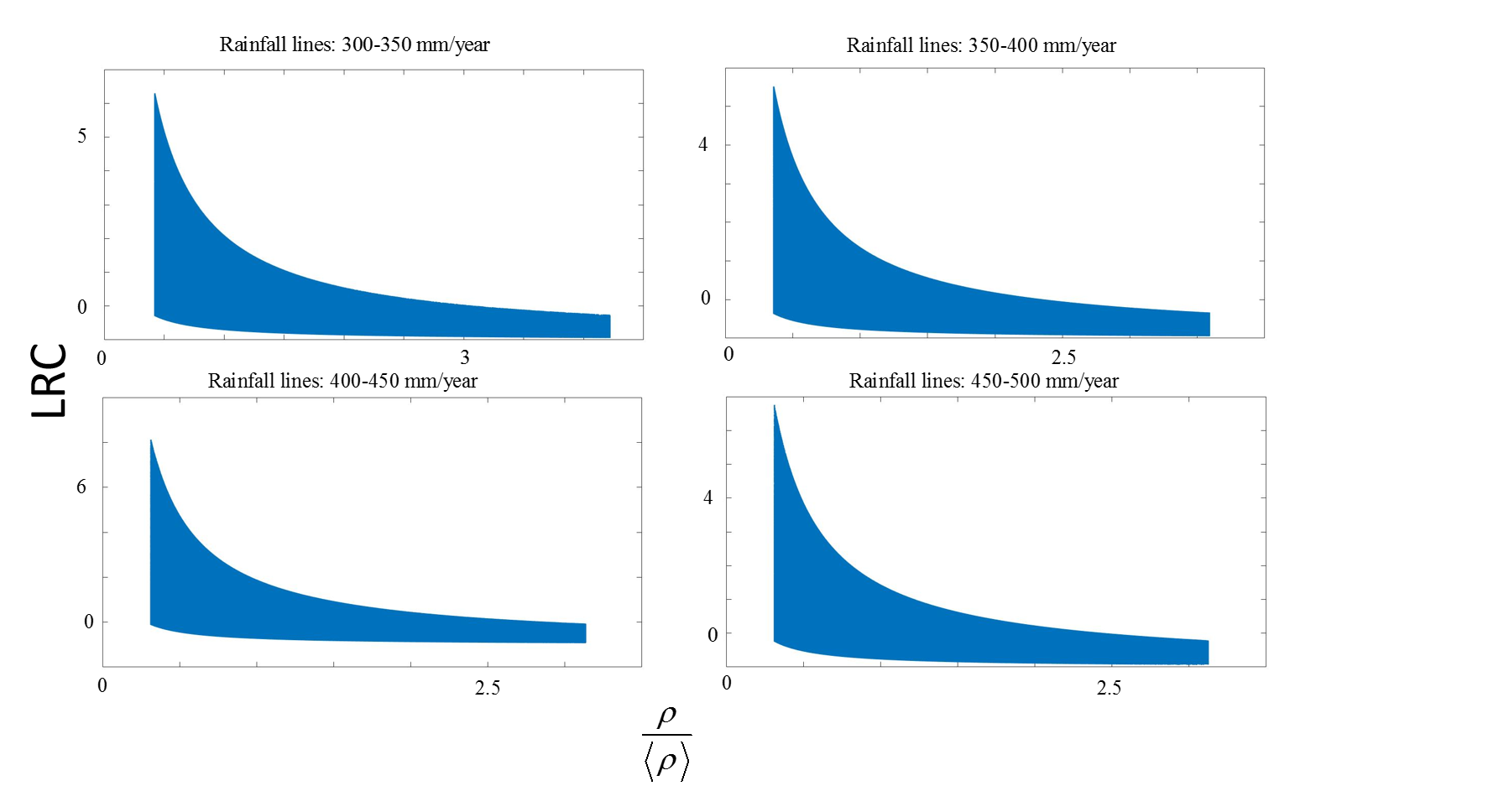}}
	\caption{\textbf{Local response curve - Data points (rainfall lines 100 - 500 [$mm/year$], years 2002-2015)}.}
\end{figure}

\begin{figure}[H]
	 \makebox[\textwidth][c]{\includegraphics[scale=0.5]{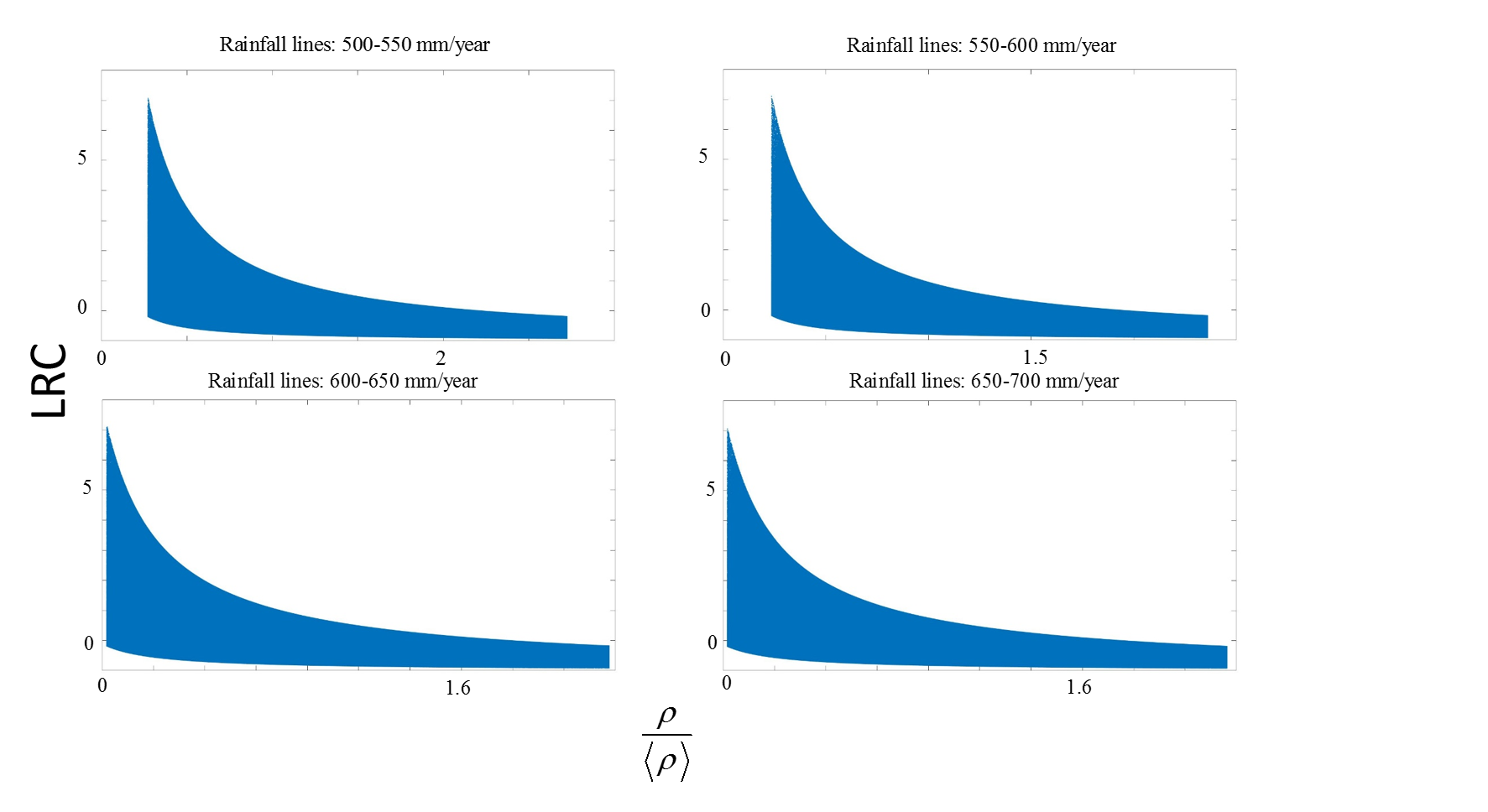}}

	\vspace{1.0pt}
	
	 \makebox[\textwidth][c]{\includegraphics[scale=0.5]{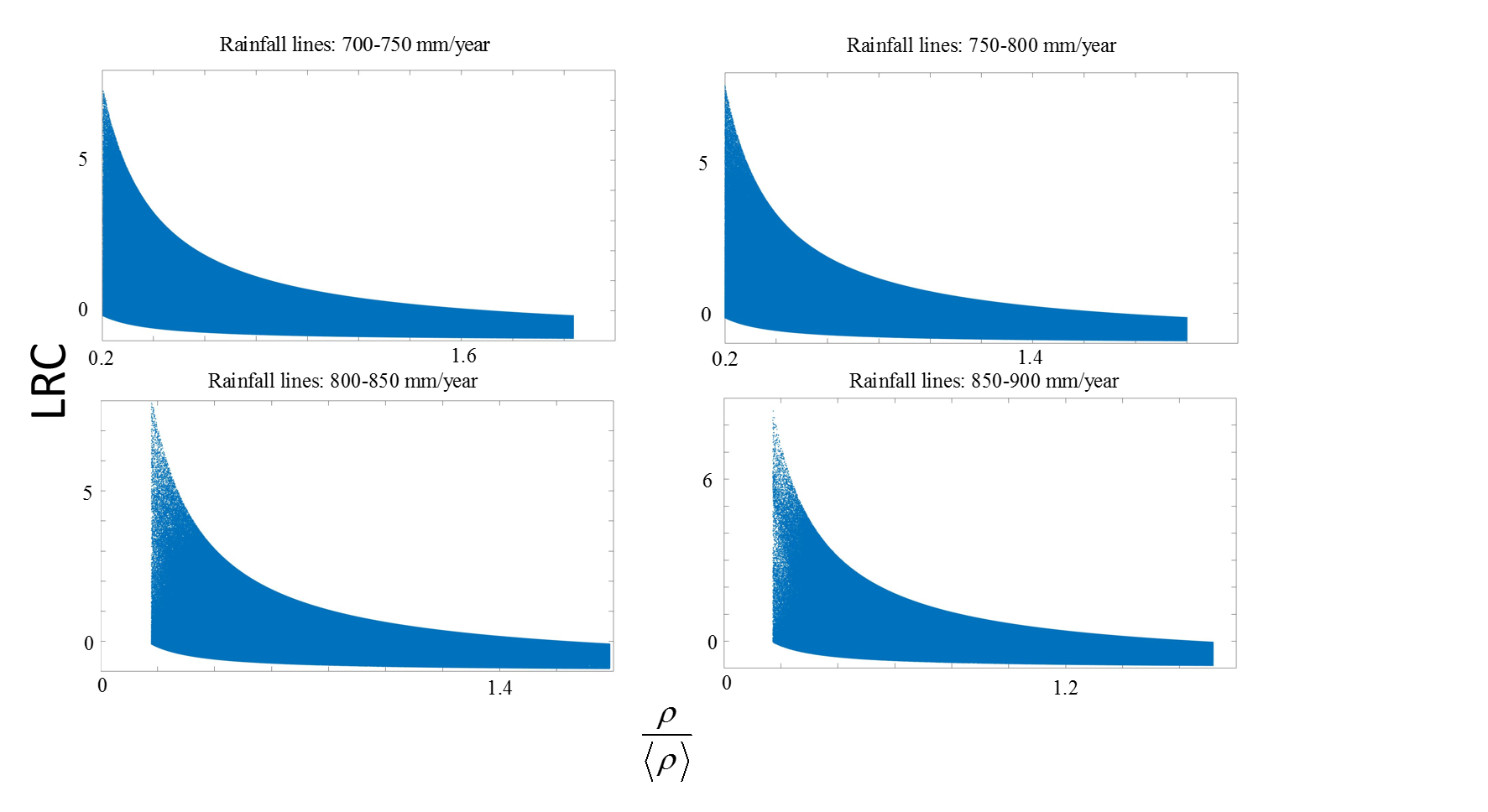}}
	\caption{\textbf{Local response curve - Data points (rainfall lines 500 - 900 [$mm/year$], years 2002-2015)}.}
\end{figure}

\newpage
\pagebreak

\begin{figure}[H]
	 \makebox[\textwidth][c]{\includegraphics[scale=0.5]{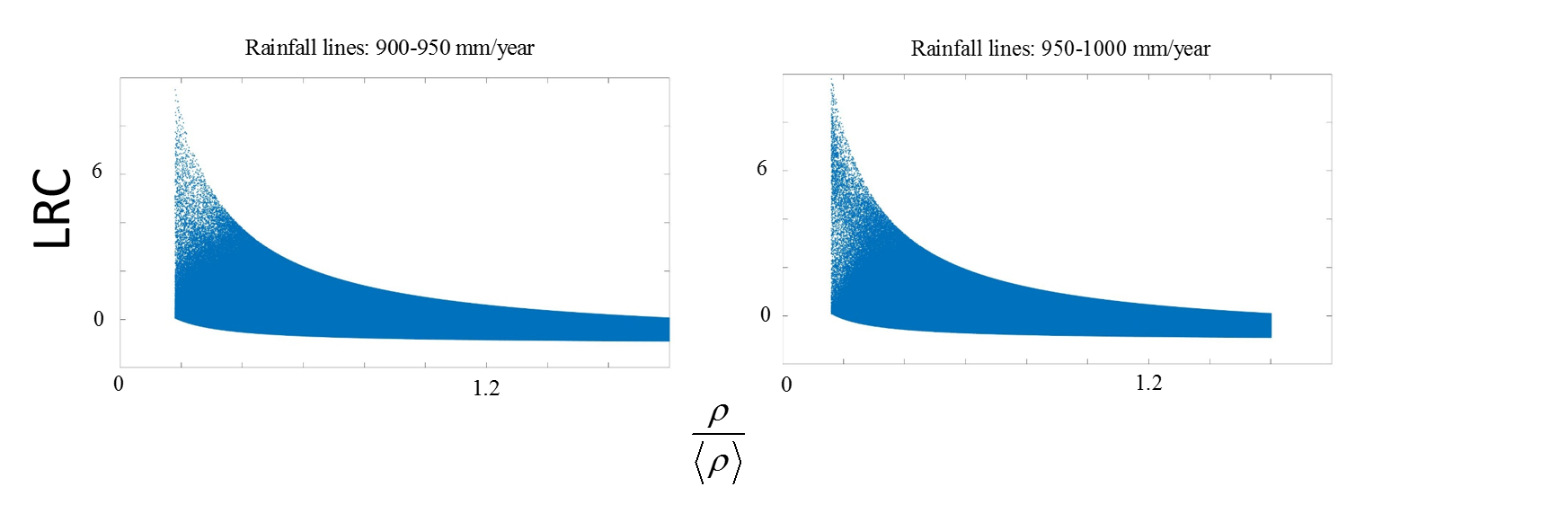}}

	\caption{\textbf{Local response curve - Data points (rainfall lines 900 - 1000 [$mm/year$], years 2002-2015)}.}
\end{figure}

\subsection*{1.4 Histograms} As a complementary for Fig. 4 of the main text, here we show the histograms of vegetation density values for different rainfall lines.
\newpage
\pagebreak
\begin{figure}[H]
	 \makebox[\textwidth][c]{\includegraphics[scale=0.5]{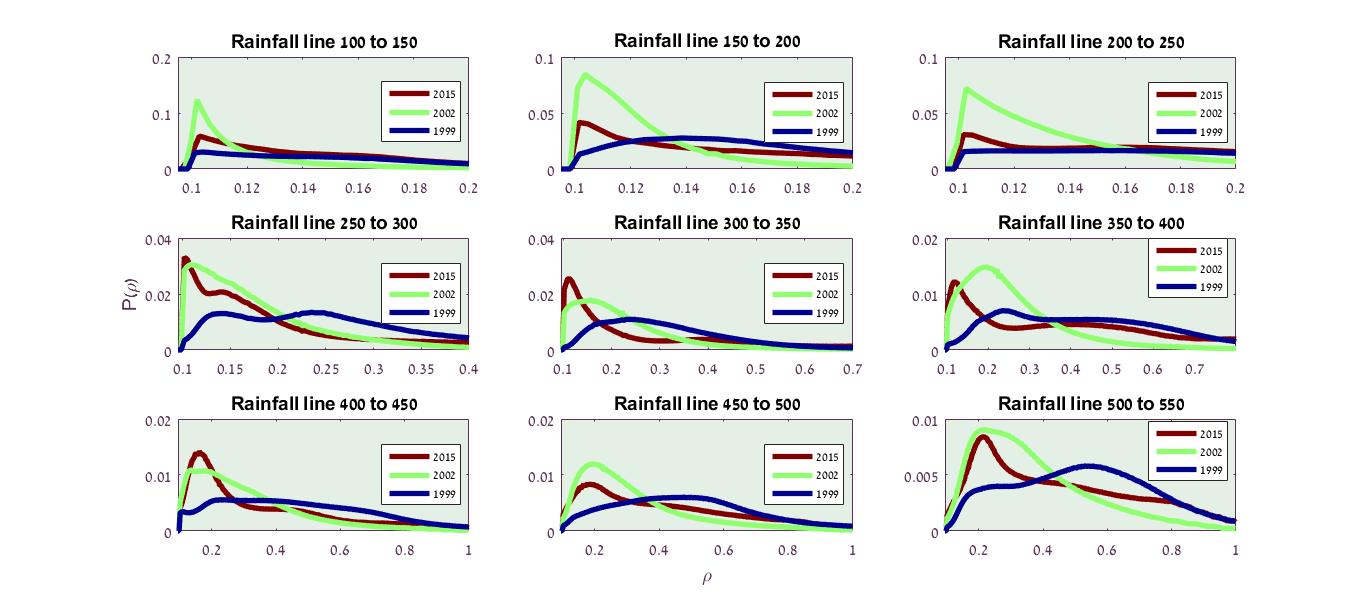}}

	\vspace{1.0pt}
	
	 \makebox[\textwidth][c]{\includegraphics[scale=0.5]{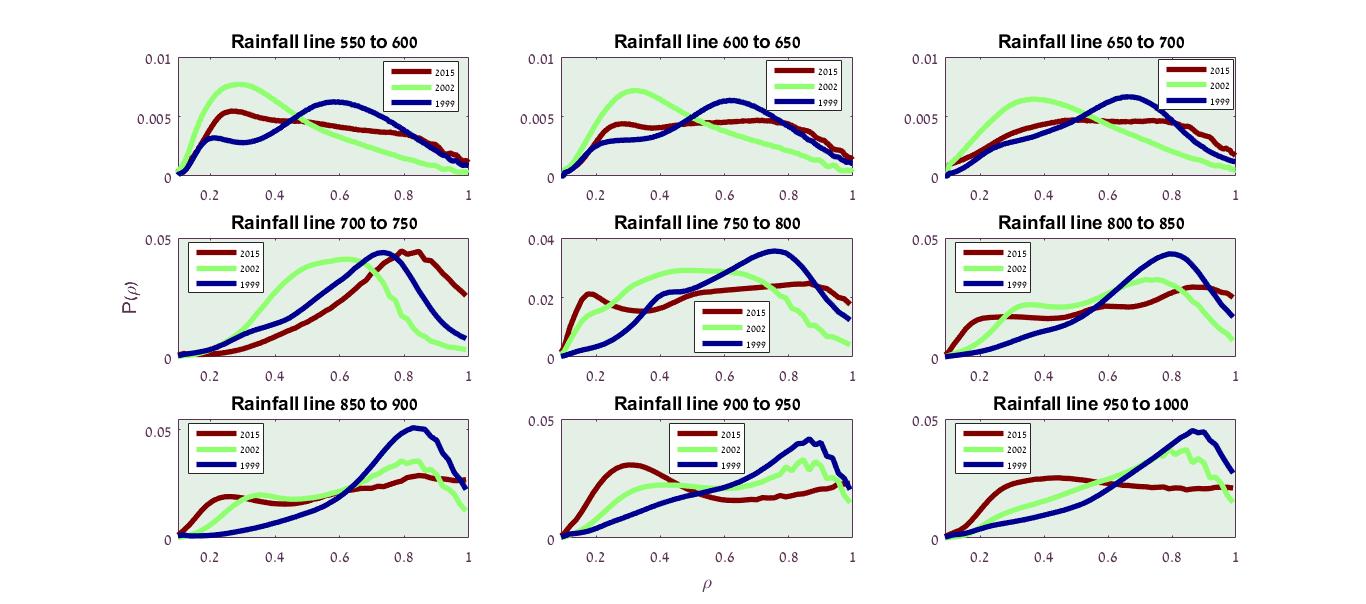}}
	\caption{\textbf{Histograms (rainfall lines 100 - 1000 [$mm/year$])}.}
\end{figure}


\end{document}